\documentclass[twocolumn,aps,superscriptaddress]{revtex4}
\usepackage{amsmath}
\usepackage{amssymb}
\usepackage{graphicx}
\usepackage{color}
\usepackage{bm}% bold math
\newcommand{\be}{\begin{equation}}
\newcommand{\ee}{\end{equation}}
\newcommand{\bea}{\begin{eqnarray}}
\newcommand{\eea}{\end{eqnarray}}
\newcommand{\bse}{\begin{subequations}}
\newcommand{\ese}{\end{subequations}}
\newcommand{\ecp}{${\rm EuCo_2P_2}$}
\newcommand{\bcp}{${\rm BaCo_2P_2}$}
\newcommand{\tcs}{${\rm ThCr_2Si_2}$}
\newcommand{\eca}{${\rm EuCo_2As_2}$}

\begin{document}

\title{EuCo$_{2}$P$_{2}$: A model molecular-field helical Heisenberg antiferromagnet}

\author{N. S. Sangeetha}
\affiliation{Ames Laboratory and Department of Physics and Astronomy, Iowa State University, Ames, Iowa 50011, USA}
\author{Eduardo Cuervo-Reyes}
\email{eduardo.cuervoreyes@empa.ch}
\affiliation{Swiss Federal Laboratories for Materials Science and Technology (Empa), \"{U}berlandstrasse 129, CH-8600 D{\"u}bendorf, Switzerland}
\affiliation{Swiss Federal Institute of Technology (ETH), Vladimir-Prelog-Weg 1, CH-8093 Z{\"u}rich, Switzerland}
\author{Abhishek Pandey}
\altaffiliation{Current address: Department of Physics and Astronomy, Texas A\&M University, College Station, Texas 77840-4242, USA}
\affiliation{Ames Laboratory and Department of Physics and Astronomy, Iowa State University, Ames, Iowa 50011, USA}
\author{D. C. Johnston}
\email{johnston@ameslab.gov}
\affiliation{Ames Laboratory and Department of Physics and Astronomy, Iowa State University, Ames, Iowa 50011, USA}

\date{\today}

\begin{abstract}

The metallic compound \ecp\  with the body-centered tetragonal \tcs\ structure containing Eu spins-7/2 was previously shown from single-crystal neutron diffraction measurements to exhibit a helical antiferromagnetic (AFM) structure below $T_{\rm N} = 66.5$~K with the helix axis along the $c$~axis and with the ordered moments aligned within the $ab$~plane.  Here we report crystallography, electrical resistivity, heat capacity, magnetization and magnetic susceptibility measurements on single crystals of this compound.  We demonstrate that \ecp\ is a model molecular-field helical Heisenberg antiferromagnet from comparisons of the anisotropic magnetic susceptibility~$\chi$, high-field magnetization and magnetic heat capacity of \ecp\ single crystals at temperature $T\leq T_{\rm N}$ with the predictions of our recent formulation of molecular field theory.  Values of the Heisenberg exchange interactions between the Eu spins are derived from the data. The low-$T$ magnetic heat capacity $\sim T^3$ arising from spin-wave excitations with no anisotropy gap is calculated and found to be comparable to the lattice heat capacity.  The density of states at the Fermi energy of \ecp\ and the related compound \bcp\ are found from the heat capacity data to be large, 10 and 16~states/eV per formula unit for \ecp\ and \bcp, respectively.  These values are enhanced by a factor of $\sim 2.5$ above those found from DFT electronic structure calculations for the two compounds.  The calculations also find ferromagnetic Eu--Eu exchange interactions within the $ab$~plane and AFM interactions between Eu spins in nearest- and next-nearest planes, in agreement with the MFT analysis of $\chi_{ab}(T\leq T_{\rm N})$.

\end{abstract}


\maketitle  

\section{\label{Sec:Intro} Introduction}

Above the transition temperature of an antiferromagnet (AFM, N\'eel temperature $T_{\rm N}$) or a ferromagnet (FM, Curie temperature $T_{\rm C}$), the magnetic susceptibility $\chi$ of a Heisenberg spin system is given within the Weiss molecular field theory (MFT) by the Curie-Weiss law \cite{Curie1895, Weiss1907}
\be
\chi \equiv \frac{M}{H} = \frac{C}{T-\theta_{\rm p}},
\label{Eq:CW-law}
\ee
where $M$ is the magnetization of the system induced in the direction of a small applied magnetic field~$H$, $C$ is the Curie constant reflecting the magnitude of the magnetic moments and $\theta_{\rm p}$ is the Weiss temperature reflecting their interactions.  Using the same MFT, the $\chi$ parallel and perpendicular to the easy axis of a collinear AFM containing identical crystallographically-equivalent spins was calculated, where the Heisenberg exchange interactions are the same between a spin and its nearest neigbors and zero otherwise \cite{Neel1936, VanVleck1941}.  This led to the unique prediction \cite{VanVleck1941} $f=-1$ for the ratio
\be
f \equiv \frac{\theta_{\rm p}}{T_{\rm N}},
\label{Eq:fRatio}
\ee
but this prediction is rarely observed quantitatively in real Heisenberg AFMs.

\begin{figure}[t]
\includegraphics [width=2.in]{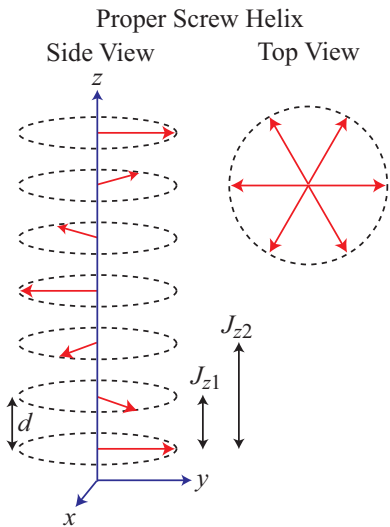}
\caption {(Color online) Generic helix AFM structure \cite{Johnston2012}.  Each arrow represents a layer of moments perpendicular to the $z$~axis that are ferromagnetically aligned within the $xy$ plane and with interlayer separation $d$.  The wave vector {\bf k} of the helix is directed along the $z$~axis.  The magnetic moment turn angle between adjacent magnetic layers is $kd$.  The exchange interactions $J_{z1}$ and $J_{z2}$ within the $J_0$-$J_{z1}$-$J_{z2}$ Heisenberg MFT model are indicated.}
\label{Fig:J0_Jz1_Jz2_model_helix}
\end{figure}

MFT calculations of the anisotropic $\chi(T\leq T_{\rm N})$ and other properties of a  noncollinear ``proper'' helix AFM structure, shown schematically in Fig.~\ref{Fig:J0_Jz1_Jz2_model_helix},  were later carried out \cite{Yoshimori1959}.  The helix is a structure in which the ordered moments that are FM-aligned within the $xy$~plane rotate their direction along the helix ($z$)~axis by a fixed angle $kd$ between adjacent planes of spins with the tips of the magnetic moment vectors tracing out the ridges on a screw.  On the other hand, when the wave vector of the helix is in the $xy$ ordering plane of the magnetic moments, the AFM structure is termed a cycloid structure \cite{Yoshimori1959}, as shown in Fig.~1 of \cite{Goetsch2014}.  The above MFT results have been little used by experimentalists to fit their $\chi(T\leq T_{\rm N})$ data for collinear, helical or cycloidal AFMs because of the difficulty of generalizing the theoretical predictions.  A comprehensive review of the theory of helical spin ordering as of 1967 is available \cite{Nagamiya1967}.

We recently reformulated the Weiss MFT for identical crystallographically-equivalent Heisenberg spins that generalizes the calculations of $\chi(T)$ and magnetic heat capacity $C_{\rm mag}(T)$ for AFMs at temperatures $T \leq T_{\rm N}$ to both collinear and coplanar noncollinear AFMs with general sets of exchange interactions including geometric and bond frustrating interactions.  This formulation can accommodate the large range of the allowed $f$ ratio in Eq.~(\ref{Eq:fRatio}) for AFMs given by $-\infty<f<1$ \cite{Johnston2012, Johnston2015}.  This formulation does not utilize the concept of magnetic sublattices usually used previously for AFMs but is instead formulated in terms of the angles in the magnetically-ordered or paramagnetic (PM) state between a central moment and its neighbors with which it interacts to calculate the thermodynamic properties.  This feature allows both collinear and noncollinear AFMs to be treated on the same footing. Another attractive feature of the MFT is that its final formulation is in terms of directly measurable quantities instead of in terms of the molecular field coupling constants or Heisenberg exchange interactions as done previously.

The prototype for a noncollinear AFM structure is the helix shown in Fig.~\ref{Fig:J0_Jz1_Jz2_model_helix}.  When the MFT was reformulated \cite{Johnston2012, Johnston2015}, there were no reported experimental anisotropic $\chi(T\leq T_{\rm N})$ data on single crystals of a material showing a helical AFM structure that satisfied the assumptions of the MFT with which to test our MFT predictions.  These assumptions are that the spin~$S$ is large to suppress quantum fluctuations, that there is no FM component to the ordering, and that there is no change in the magnetic structure below $T_{\rm N}$.  However, a 1992 neutron diffraction study of the body-centered tetragonal compound ${\rm EuCo_2P_2}$ found that the Eu spins $S = 7/2$ exhibit helical ordering with no change in the magnetic structure from $T_{\rm N}$ down to at least 15~K \cite{Reehuis1992}. This result motivated us to grow crystals of \ecp\ and measure their properties to test the applicability of our MFT to a helical AFM.

\ecp\ has the \tcs{} structure with space group $I4/mmm$ \cite{Marchand1978}.  Magnetic susceptibility $\chi$ measurements versus temperature~$T$ of a polycrystalline sample \cite{Morsen1988} as well as the neutron diffraction measurements on a single crystal \cite{Reehuis1992} and other measurements \cite{Nakama2010} demonstrate AFM ordering of the Eu$^{+2}$ spins $S=7/2$ at $T_{\rm N} = 66.5(5)$~K with no contribution from the Co atoms.  The $\chi(T>T_{\rm N})$ follows the Curie-Weiss law~(\ref{Eq:CW-law}) with $\theta_{\rm p} = 20(2)$~K indicating dominant ferromagnetic interactions \cite{Morsen1988}.

The ordered moment of \ecp\ at 15~K from the neutron diffraction study is $\langle\mu\rangle = 6.9(1)~\mu_{\rm B}$/Eu \cite{Reehuis1992}, which agrees with the saturation moment $\mu_{\rm sat} = gS\mu_{\rm B}$/Eu $= 7\,\mu_{\rm B}$/Eu expected for Eu spin $S=7/2$ and $g=2$.  Here $\mu_{\rm B}$ is the Bohr magneton and $g$ is the spectroscopic splitting factor.  The authors discovered that the magnetic structure is a planar helix with the Eu ordered moments aligned in the $ab$~plane of the tetragonal structure, with the helix axis along the perpendicular $c$~axis.  The observed $ab$-plane alignment of the ordered moments is consistent with the prediction of the moment alignment from magnetic dipole interactions between the Eu spins \cite{Johnston2016}.  The incommensurate AFM propagation vector changed by 2.1\% from ${\bf k} = [0,\ 0,\ 0.834(4)]2\pi/c$ at $T=64$~K to $[0,\ 0,\ 0.852(4)]2\pi/c$ at $T=15$~K, where $c$ is the $c$-axis latttice parameter of the body-centered tetragonal Eu sublattice.  Since $d=c/2$ is the distance along the helix $c$~axis between adjacent layers of FM-aligned moments, the turn angle $kd$ between the ordered moments in adjacent layers is
\be
kd(64~{\rm K}) = 0.834(4)\pi,\quad  kd(15~{\rm K}) = 0.852(4)\pi.
\label{Eq:kdNeuts}
\ee
These values with $\pi/2 < kd < \pi$ indicate that the dominant {\it interlayer} interactions are AFM  \cite{Johnston2012, Johnston2015}, and the above $\theta_{\rm p} = 20(2)$~K \cite{Morsen1988} together with Eq.~(\ref{eq:thetap}) below therefore indicate that the dominant {\it intralayer} interactions must be FM\@.  The radius of nonmagnetic Sr$^{+2}$ and magnetic Eu$^{+2}$ are similar, and it is important in the present context to note that neither ${\rm SrCo_2P_2}$ \cite{Morsen1988, Sugiyama2014} nor ${\rm SrCo_2As_2}$ \cite{Pandey2013} exhibit long-range magnetic order.

Experiments on \ecp\ at high pressure reveal a first-order tetragonal to collapsed-tetragonal \cite{Anand2012}  transition and associated valence transition from Eu$^{+2}$ to nonmagnetic Eu$^{+3}$ at $\sim 3$~GPa \cite{Huhnt1997, Chefki1998}, together with a change from Eu(4$f$) local moment to Co(3$d$) itinerant magnetic ordering \cite{Chefki1998}.  In the isostructural compound ${\rm EuCo_2As_2}$, a continuous tetragonal to collapsed-tetragonal transition occurs at a pressure of $\approx 5$~GPa \cite{Bishop2010}, whereas in ${\rm SrCo_2As_2}$ a first-order tetragonal to collapsed-tetragonal transition is observed at about 6~GPa \cite{Jayasekara2015}.

Herein, we report room-temperature crystallography results for crushed \ecp{} single crystals together with electrical resistivity~$\rho$, heat capacity~$C_{\rm p}$, $M$ and~$\chi$ measurements versus $T$ and~$H$ for single crystals.  We analyze the magnetic data in terms of the MFT predictions for a helical AFM structure.  We conclude from these results that \ecp\ is a model molecular-field helical Heisenberg antiferromagnet.  {\it Ab~initio} electronic structure calculations in the generalized gradient (GGA) approximation are presented that support this conclusion.  

The experimental details are given in Sec.~\ref{Sec:ExpDetails}.  The structural refinement of \ecp\ is presented in Sec.~\ref{Sec:Struct} which confirms previous results \cite{Marchand1978}.  The in-plane $\rho(H,T)$ data are presented in Sec.~\ref{Sec:Rho} and the $C_{\rm p}(T,H)$ results in Sec.~\ref{Sec:Cp}.  The $M(H)$ isotherm and $\chi(T)$ data are presented in Sec.~\ref{Sec:MChi}.  The Heisenberg exchange interactions in \ecp\ are estimated in Sec.~\ref{Sec:HeisExchInts}.  We also obtain an estimate of the classical ground-state energy of the helix.  In Sec.~\ref{Sec:SpinWaves} we study the spin-wave spectrum in the absence of an anisotropy gap and determine the low-energy anisotropic spin-wave velocities in the helix.  We then calculate the $T^3$ spin-wave contribution to the low-temperature $C_{\rm p}(T)$ and find that it is comparable to the lattice contribution.  The electronic structure calculations are presented in Sec.~\ref{Sec:ElecStruct}.  A summary of our results and conclusions is given in Sec.~\ref{Sec:Summary}.

\section{\label{Sec:ExpDetails} Experimental Details}

Single crystals of \ecp\ were grown in Sn flux as described previously \cite{Reehuis1992} whereas a polycrystalline sample of \bcp\ was prepared by solid state reaction. Rietveld refinement of powder x-ray diffraction (XRD) data for \bcp\ with the ${\rm ThCr_2Si_2}$ structure yielded $a=3.8057(2)$\,\AA, $c = 12.4115(5)$\,\AA\ and $z_{\rm P} = 0.3565(4)$.  Semiquantitative chemical analysis of the \ecp\ crystals was performed using a JEOL scanning electron microscope (SEM), equipped with an EDX (electron dispersive x-ray spectroscopy) analyzer. The EDX measurements showed the expected 1:2:2 stoichiometry. Room-temperature powder XRD measurements with Cu $\rm{K_{\alpha}}$ radiation were carried out on \ecp\ crushed crystals with a Rigaku Geigerflex x-ray diffractometer. The data were analysed by Rietveld refinement using FullProf software \cite{fullprof}.  Magnetization data were obtained using a Quantum Design, Inc., magnetic properties measurement system (MPMS) and a vibrating sample magnetometer in a Quantum Design, Inc., physical properties measurement system (PPMS) for high-field measurements up to 14~T, where 1~T~$\equiv10^{4}$~Oe. A PPMS was also used for $C_{{\rm p}}(T)$ and $\rho(T)$ measurements. The $C_{{\rm p}}(T)$ was measured by the relaxation method and the $\rho(T)$ was measured using the standared four-probe ac technique.

\section{\label{Sec:Struct} Crystallography}

\begin{figure}
\includegraphics[width=3.5in]{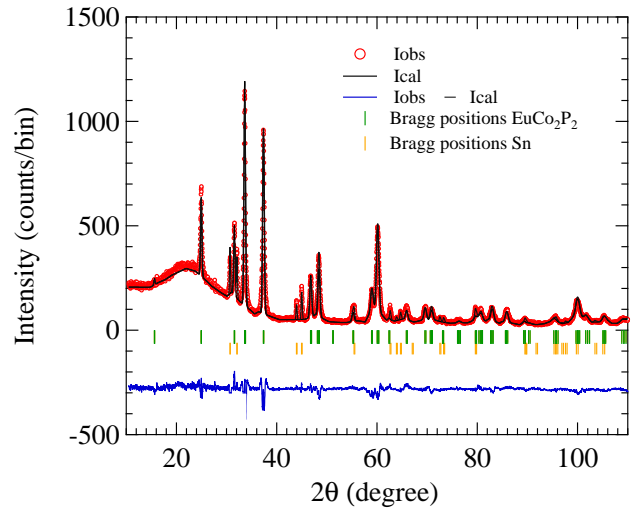}
\caption{(Color online) Powder XRD pattern of \ecp{} at 300 K\@. The solid line through the experimental points is the two-phase Rietveld refinement fit calculated for the \tcs-type crystal structure (space group $I4/mmm$) including the bct $\beta$-Sn impurity phase structure (space group $I4_{1}/amd$).}
\label{fig1:EuCo2P2_xrd}
\end{figure}

The powder x-ray diffraction (XRD) pattern of \ecp{} at room temperature is shown in Fig.~\ref{fig1:EuCo2P2_xrd}. The Rietveld refinement confirms that \ecp{} has the \tcs-type crystal structure with space group $I4/mmm$. The crystal structure data and refinement parameters obtained are summarized in Table~\ref{Table:EuCo2P2_xrd parameter}.  The lattice parameters are in good agreement with previously-reported values \cite{Marchand1978}.  The XRD pattern also reveals the presence of metallic Sn impurity that arises from a small amount of adventitious Sn flux on the surfaces and/or embedded in the sample which is accounted for using the two-phase refinement in Fig.~\ref{fig1:EuCo2P2_xrd}.  

\begin{table}
\caption{\label{Table:EuCo2P2_xrd parameter} Crystal and refinement parameters and atomic coordinates obtained from Rietveld refinement of room-temperature powder XRD data of crushed \ecp{} crystals with the \tcs-type crystal structure, space group $I4/mmm$ and $Z=2$ formula units per unit cell.}
\begin{ruledtabular}
\begin{tabular}{lcccc}
Atom 	& Wyckoff position	& $x$	& $y$  	& $z$		\\
\hline 
Eu  		& 2$\mathit{a}$  	& 0  	& 0  	& 0			\\
Co  		& 4$\mathit{d}$  	& 0  	& 1/2  	& 1/4		\\
P  		& 4$\mathit{e}$  	& 0  	& 0  	& 0.3558(5)	\\
\hline 
Lattice parameters  &  &  &  & \\
$a$~(\AA)					& 3.7597(3)	\\
$c$~(\AA)					& 11.3369(4)	\\
$c/a$  					& 3.015(3)	\\
$V_{\rm cell}~(\rm{\AA}^3)$	& 160.25(3)	\\
\hline
Rietveld fit parameters				\\
$\chi^2$					& 3.15		\\
$R_{\rm p}$ (\%)			& 11.9		\\
$R_{\rm wp}$ (\%)			& 16.4		\\
\end{tabular}
\end{ruledtabular}
\end{table}

\section{\label{Sec:Rho} Electrical Resistivity}

\begin{figure}
\includegraphics[width=3.5in]{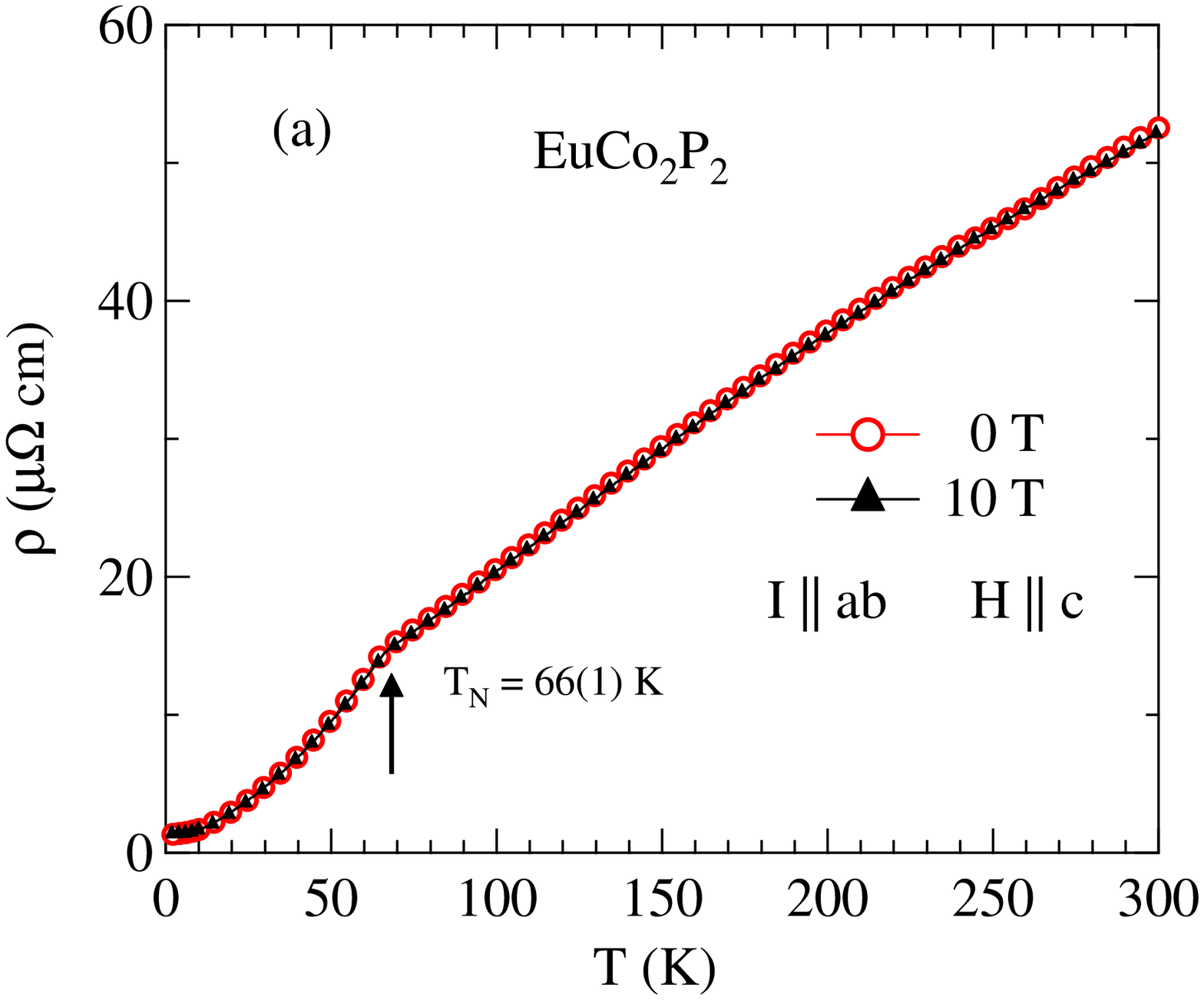}
\includegraphics[width=3.5in]{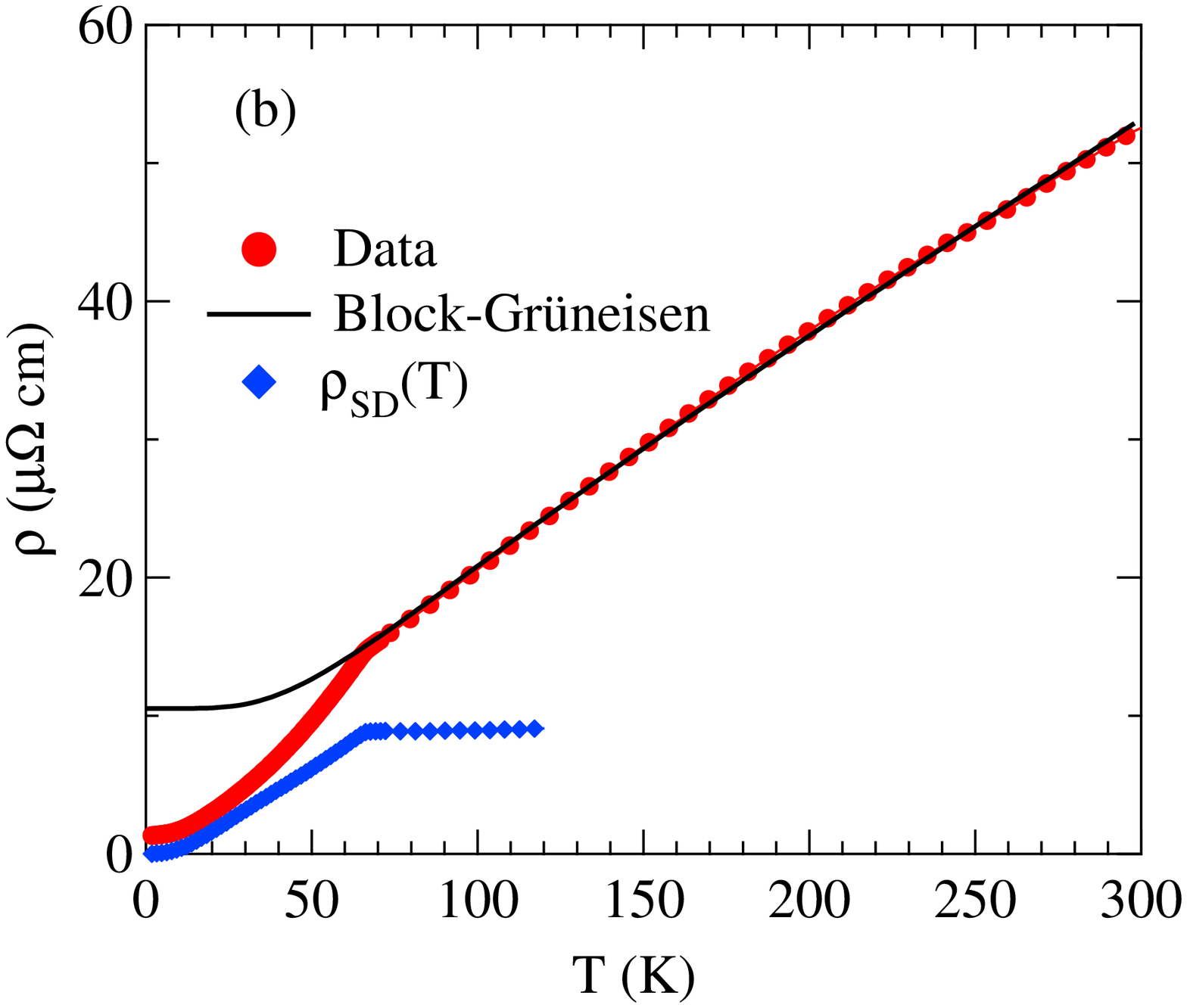}
\protect\caption{(a) In-plane electrical resistivity $\rho(T)$ of \ecp{} in the temperature~$T$ range 1.8 to 300~K for $H=0$~T and 10~T applied parallel to the $c$ axis. The vertical arrow indicates the AFM transition. (b)~$\rho(T)$ of \ecp{} in $H=0$. The curve is the fit of Eqs.~(\ref{Eqs:RhoBG}) to the data above 70~K and is extrapolated to $T=0$.  The spin-disorder scattering resistivity $\rho_{\rm SD}(T)$ plotted as filled diamonds is obtained from Eqs.~(\ref{Eqs:RhoSD(T)}). }
\label{Fig:EuCo2P2_Rho}
\end{figure}

The $ab$-plane $\rho(T)$ of an \ecp{} crystal at $H = 0$~T and 10~T measured in the temperature range 1.8 to 300~K are shown in Fig.~\ref{Fig:EuCo2P2_Rho}(a). The $\rho(T)$ shows a metallic behaviour with a residual resistivity ratio RRR~$\equiv \rho$(300~K)/$\rho$(2\,K) = 39.4. The large RRR and the small value of $\rho(2~{\rm K) = 1.3\,\mu\Omega\,cm}$ indicate that the crystals are of high quality.  The $\rho(T)$ shows a sudden increase of slope upon cooling below $T_{\rm N} = 66(1)$~K, indicated by an arrow in Fig.~\ref{Fig:EuCo2P2_Rho}(a), which we ascribe below to a reduction in spin-disorder (SD) scattering.  Our $\rho(T)$ data are very similar to those presented previously for a single crystal of \ecp \cite{Nakama2010}.  The $T_{\rm N}$ determined from our $\rho(T)$ data agrees with that found from our $C_{\rm p}(T)$ and $\chi(T)$ data below and with literature values \cite{Morsen1988,Nakama2010,Reehuis1992}.  From Fig.~\ref{Fig:EuCo2P2_Rho}(a)  the magnetoresistance is negligible from 2~K to 300~K in a field of 10~T\@.

The $\rho(T)$ data in $H= 0$ are shown separately in Fig.~\ref{Fig:EuCo2P2_Rho}(b).  We assume the validity of Matthiessen's rule \cite{Ziman1960, Blatt1968, Ashcroft1976} and three contributions to $\rho$: (i) $\rho_0$ due to $T$-independent impurity scattering, (ii)~$\rho_{\rm BG}(T)$ due to electron-phonon scattering using the Bloch-Gr\"uneisen prediction, and (iii)~$\rho_{\rm SD}(T)$ due to SD scattering.

In the high-$T$ regime with $T\geq T_{\rm N}$ where $\rho_{\rm SD}$ is assumed to be a constant $\equiv \rho_{\rm SD0}$, we have
\bse
\label{Eqs:RhoBG}
\be
\rho(T\geq T_{\rm N}) = \rho_0 + \rho_{\rm SD0} + \rho_{\rm BG}(T),
\ee
where \cite{Ziman1960, Blatt1968, Goetsch2012}
\be
\rho_{\rm BG}(T) = F\left(\frac{T}{\Theta_{\rm R}}\right)^{5}\int_{0}^{\Theta_{\rm R}/T}\frac{x^{5}dx}{\left(1-e^{-x}\right)\left(e^{x}-1\right)},
\label{Eq:FigRhoHighT}
\ee
$F$ is a numerical constant that describes the $T$-independent interaction strength of the conduction electrons with the thermally excited phonons and contains the average atomic mass and conduction carrier Fermi velocity, and $\Theta_{\rm R}$ is the resistively-determined Debye temperature.  The representation for $\rho_{\rm BG}(T)$ used here is an accurate analytic Pad\'e approximant function of $T/\Theta_{\rm R}$ \cite{Goetsch2012}.  We fitted the data for $T\geq T_{\rm N}$ (70~K~$\leq T \leq 340$~K) by 
Eq.~(\ref{Eq:FigRhoHighT}) as shown the black curve in Fig.~\ref{Fig:EuCo2P2_Rho}(b), where an extrapolation of the fit to $T=0$ is shown for which the $y$~intercept is $\rho_0+\rho_{\rm SD0}$ since $\rho_{\rm BG}(T=0)=0$.  The parameters found from the fit are
\be
\rho_0 + \rho_{\rm SD0} = 10.5(7)~{\rm \mu\Omega\,cm},\quad \Theta_{\rm R} = 264(6)~{\rm K}.
\ee
\ese

In the low-$T$ range with $0\leq T\leq T_{\rm N}$, we first carried out a quadratic fit to the $\rho(T)$ data from 1.8 to 15~K and obtained
\bse
\label{Eqs:RhoSD(T)}
\be
\rho_0 = 1.3(1)~{\rm \mu\Omega\,cm}.
\ee
Then the spin-disorder scattering is obtained from
\be
\rho_{\rm SD}(T) = \rho(T) - \rho_0 - \rho_{\rm BG}(T).
\label{Eq:rhoSD(T)}
\ee
\ese
A plot of the $\rho_{\rm SD}$ versus~$T$ data is shown as the filled diamonds in Fig.~\ref{Fig:EuCo2P2_Rho}(b).  The data below~$T_{\rm N}$ do not follow a single power law.

\section{\label{Sec:Cp} Heat Capacity}

\begin{figure}
\includegraphics[width=3.5in]{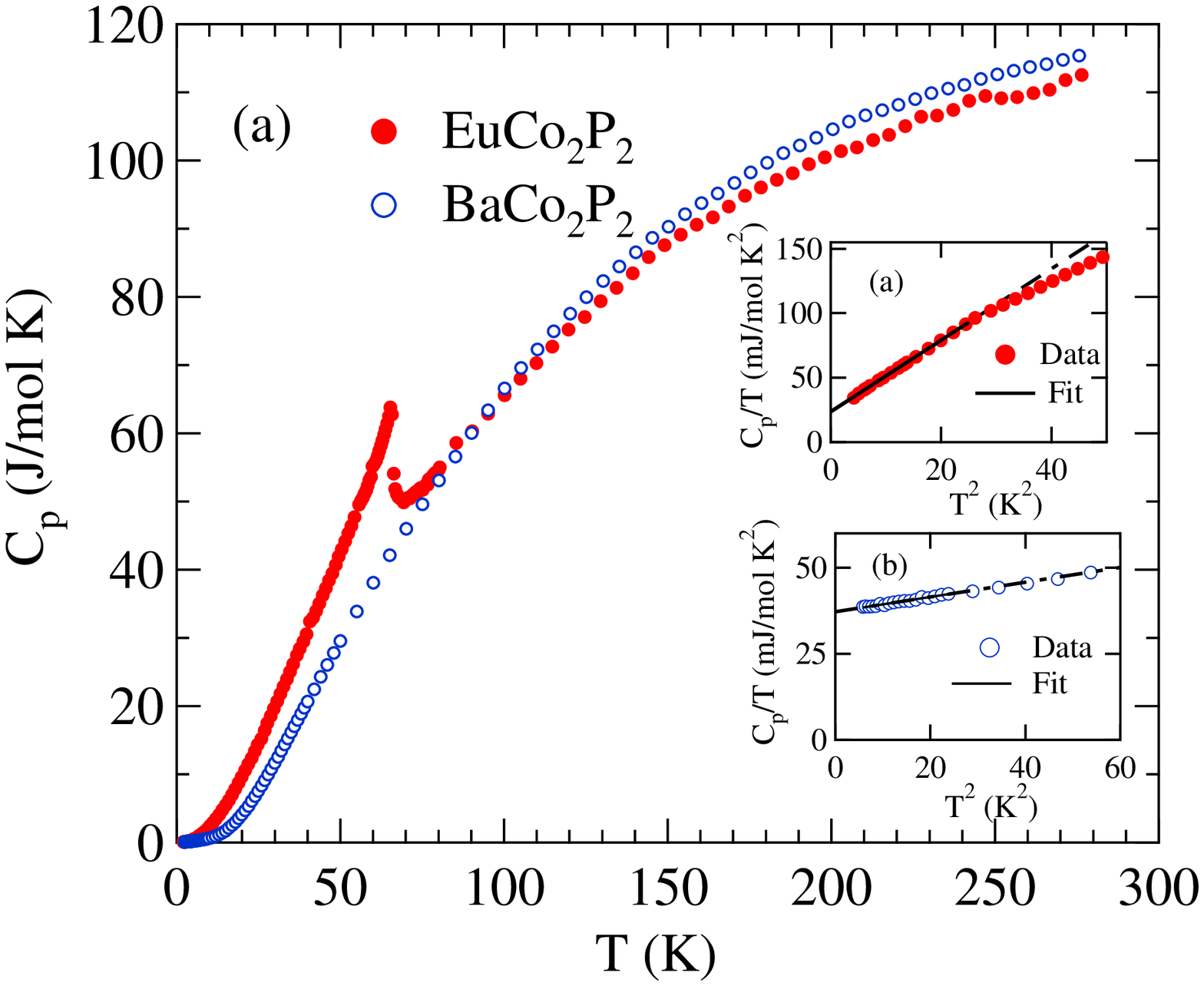}
\includegraphics[width=3.5in]{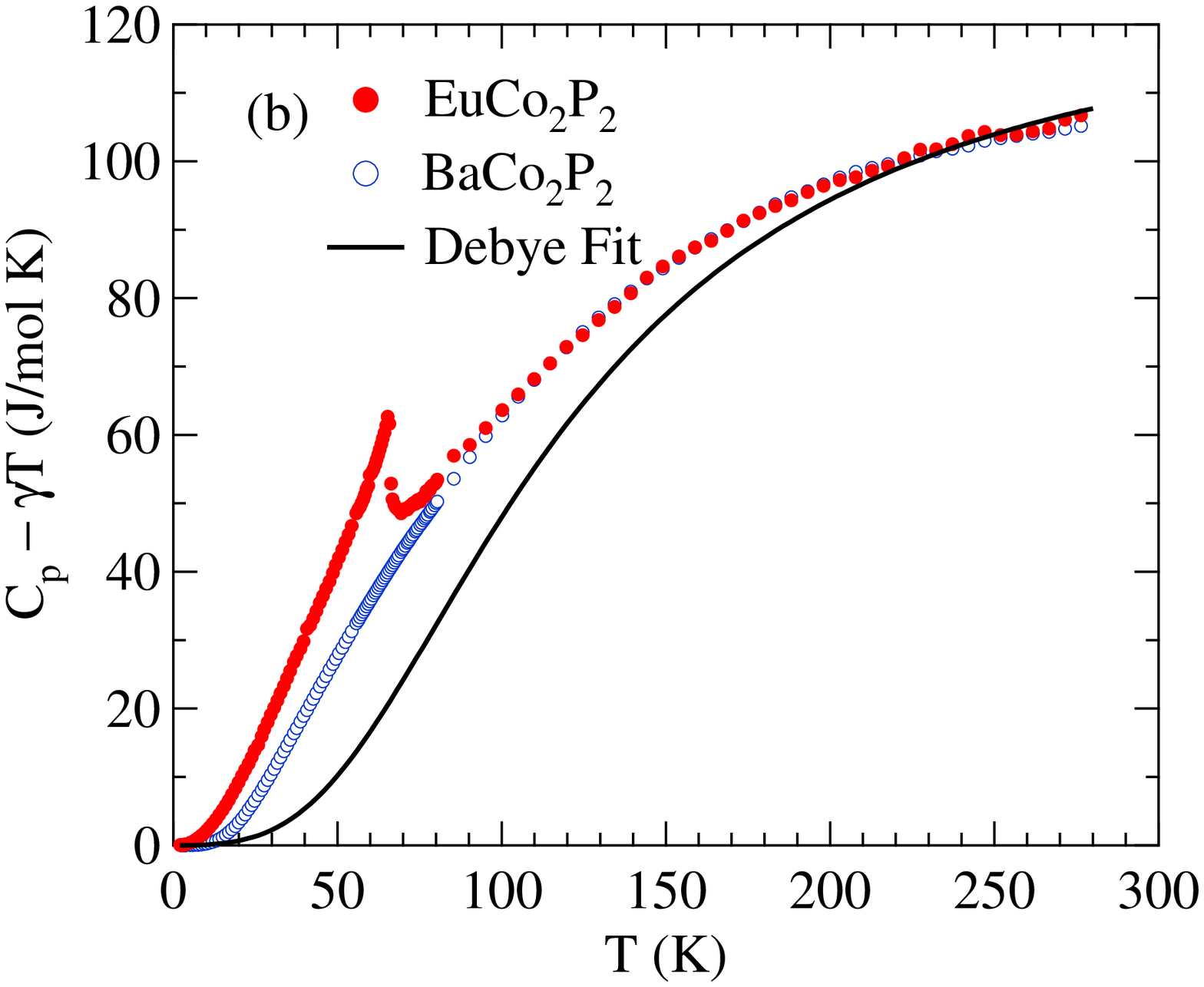}
\caption{(Color online) (a) Temperature dependence of the heat capacity $C_{\rm p}$ for  \ecp{} and \bcp{} in $H=0$. Insets~(1) and~(2): $C_{\rm p}/T$ versus $T^2$ for \ecp{} and \bcp{}, respectively, where the straight lines are fits of the data between 1.8~K and 5~K by Eq.~(\ref{Eq:Cp_Fit}). (b)~$C_{\rm p}-\gamma T$ versus $T$ for \ecp{} and \bcp{}.  The black curve is a fit of the data between 200 and 280~K by the Debye lattice heat capacity model in Eqs.~(\ref{Eqs:HiTFit}).}
\label{Fig: EuCo2P2_Cp}
\end{figure}

The $C{\rm_p}(T)$ data for \ecp{} and for the reference compound \bcp\ measured in the temperature range from 1.8 to 280~K are shown in Fig.~\ref{Fig: EuCo2P2_Cp}(a). A sharp peak is seen in $C_{\rm p}(T)$ of \ecp\ at $T_{\rm N} = 65.7(1)$~K\@. The $C_{\rm p}(T)$ of \bcp{} is typical of a nonmagnetic metallic material. The low-$T$ $C_{\rm p}(T)$ data for \ecp{} and \bcp{} in the insets of Fig.~\ref{Fig: EuCo2P2_Cp}(a) were fitted over the temperature range from 1.8 to 5~K by the expression \cite{Kittel2005}
\bea
\frac{C_{{\rm p}}(T)}{T}=\gamma + \beta{T^2},
\label{Eq:Cp_Fit}
\eea
where $\gamma T$ is the electronic contribution and $\beta T^3$ contains the lattice contribution to $C_{\rm p}(T)$. 
The fits are shown by straight lines in Fig.~\ref{Fig: EuCo2P2_Cp}(a)~insets~(1) and~(2) for \ecp {} and \bcp{} respectively, and the fitting parameters $\gamma$ and $\beta$ for each compound are listed in Table.~\ref{Table:Heat_capacity}. 

The density of conduction carrier states at the Fermi energy $E_{\rm F}$ for both spin directions as measured by $C_{\rm p}$, ${\cal D}_\gamma(E_{\rm F})$, is obtained from $\gamma$ according to \cite{Kittel2005}
\bse
\be
{\cal D}_\gamma(E_{\rm F}) = \frac{3\gamma}{\pi^2k_{\rm B}^2},
\ee
which gives
\be
{\cal D}_\gamma(E_{\rm F})\left[{\rm \frac{states}{eV\,f.u.}}\right] = \frac{1}{2.359}\gamma\left[{\rm \frac{mJ}{mol\,K^2}}\right].
\label{Eq:DofEF}
\ee
\ese
The ${\cal D}_\gamma(E_{\rm F})$ values calculated for \ecp\ and \bcp\ from the $\gamma$ values in Table~\ref{Table:Heat_capacity} using  Eq.~(\ref{Eq:DofEF}) are listed in Table~\ref{Table:Heat_capacity}.  These values are large even for transition metals.  A calculation of ${\cal D}_\gamma(E_{\rm F})$ for isoelectronic nonmagnetic ${\rm SrCo_2As_2}$ was carried out using density functional theory, yielding ${\cal D}_\gamma(E_{\rm F}) = 4.0$~states/eV\,f.u.\ for both spin directions \cite{Cuervo-Reyes2014}, which is large but only 40\% of our measured value.

\begin{table}
\caption{\label{Table:Heat_capacity} Parameters $\gamma$ and $\beta$ obtained by fitting the zero-field $C_{{\rm p}}(T)$ data of \ecp\ and \bcp\ in the temperature range 1.8 to 5~K by Eq.~(\ref{Eq:Cp_Fit}). Also listed are the Debye temperature $\Theta_{\rm D}$ obtained from $\beta$ according to Eq.~(\ref{Eq:thetaD}) and the density of states at the Fermi energy ${\cal D}_\gamma(E_{\rm F})$ obtained from $\gamma$ via Eq.~(\ref{Eq:DofEF}). Another value of $\Theta_{\rm D}$ for both \ecp\ and \bcp\ is obtained by fitting the $C_{\rm p}-\gamma T$ data between 200 and 280~K in Fig.~\ref{Fig: EuCo2P2_Cp}(b) by the Debye model according to Eqs.~(\ref{Eqs:HiTFit}). }
\begin{ruledtabular}
\begin{tabular}{ccccc}
        		& $\gamma$       	& $\beta$        		& $\Theta_{\rm D}$		&	${\cal D}_\gamma(E_{\rm F})$  \\
 compound      & (mJ/mol K$^2$)  	& (mJ/mol K$^4$)		&         (K)    		&	$\left({\rm \frac{states}{eV\,f.u.}}\right)$ \\
\hline 
\ecp\  		& 23.7(5)       	&  2.8(1)\footnotemark[1]  	& 151(2)\footnotemark[1]	&	10.0(2)  	\\
			&				&   					& 480(6)\footnotemark[2]			\\
\bcp\ 		& 37.3(3)      	& 0.21(1)     			& 359(6)        			&	15.8(2)\\
\end{tabular}
\end{ruledtabular}
\footnotetext[1]{The $\beta$ value is too large to come completely from lattice vibrations.  We infer that there is a large contribution to $\beta$ from AFM spin waves of the helix, and hence the derived $\Theta_{\rm D}$ is too small.  See Sec.~\ref{Sec:SpinWaves}.}
\footnotetext[2]{200--280~K fit by Eqs.~(\ref{Eqs:HiTFit}).}
\end{table}

From the values of $\beta$, we estimate the Debye temperatures $\Theta_{\rm D}$ for the two compounds from the expression \cite{Kittel2005} 
\be
\Theta_{{\rm D}} = \left(\frac{12\pi^{4\,} R\, n\,}{5\beta}\right)^{1/3},
\label{Eq:thetaD}
\ee
where $R$ is the molar gas constant and $n = 5$ is the number of atoms per formula unit (f.u.).  The value $\Theta_{\rm D} = 151(1)$~K for \ecp{} is much smaller than the value of 359(2)~K for isostructural \bcp.  The value for \ecp\ is also much smaller than the value of 348~K obtained previously for the similar compound ${\rm SrNi_2P_2}$ \cite{Ronning2009}.  Therefore it is likely that the $\beta$ value measured for \ecp\ contains a significant contribution from three-dimensional AFM spin waves associated with the ordered Eu moments which also give a $T^3$ contribution to $C_{\rm p}$.  This inference is supported by the spin-wave calculations in Sec.~\ref{Sec:SpinWaves}. 

We used the Debye model for the lattice heat capacity to fit the $C_{\rm p}(T)-\gamma T$ data for \ecp\ and \bcp\ in Fig.~\ref{Fig: EuCo2P2_Cp}(b) by
\bse
\label{Eqs:HiTFit}
\be
C_{\rm p}- \gamma T = n C_{\rm V\,Debye},
\ee
where $C_{\rm V\,Debye}$ is the Debye lattice heat capacity per mole of atoms given by \cite{Kittel2005}
\be
C_{\rm V\,Debye} = 9 R\left(\frac{T}{\Theta_{\rm D}}\right)^{3}\int_{0}^{\Theta_{\rm D}/T}\frac{x^{4}e^x}{\left(e^x-1\right)^2 }dx.
\label{Eq:Debye}
\ee
\ese
The representation of the Debye function used here is an accurate analytic Pad\'e approximant function of $T/\Theta_{\rm D}$ \cite{Goetsch2012}.  The fit of the  $C_{\rm p}(T)-\gamma T$ data for \ecp\ and \bcp\ in Fig.~\ref{Fig: EuCo2P2_Cp}(b) by Eqs.~(\ref{Eqs:HiTFit}) over the temperature range 200 to 280~K is shown by the black curve with the fitting parameter $\Theta_{\rm D} = 480(6)$~K as listed in Table~\ref{Table:Heat_capacity}.  This value of $\Theta_{\rm D}$ is much larger than the value of 359~K obtained from the fit to the $C_{\rm p}$ data for \bcp\ at low~$T$\@.  Furthermore, the fit of the Debye model to the data in Fig.~\ref{Fig: EuCo2P2_Cp}(b) is poor below 200~K, suggesting that the density of phonon states versus energy in the Debye model ($\propto \omega^2$) is a poor approximation to those in \ecp\ and \bcp.  It is also conceivable that the Co atoms have a significant magnetic contribution to $C_{\rm p}(T)$ even though they show no long-range magnetic order in our temperature range.

\begin{figure}
\includegraphics[width=3.3in]{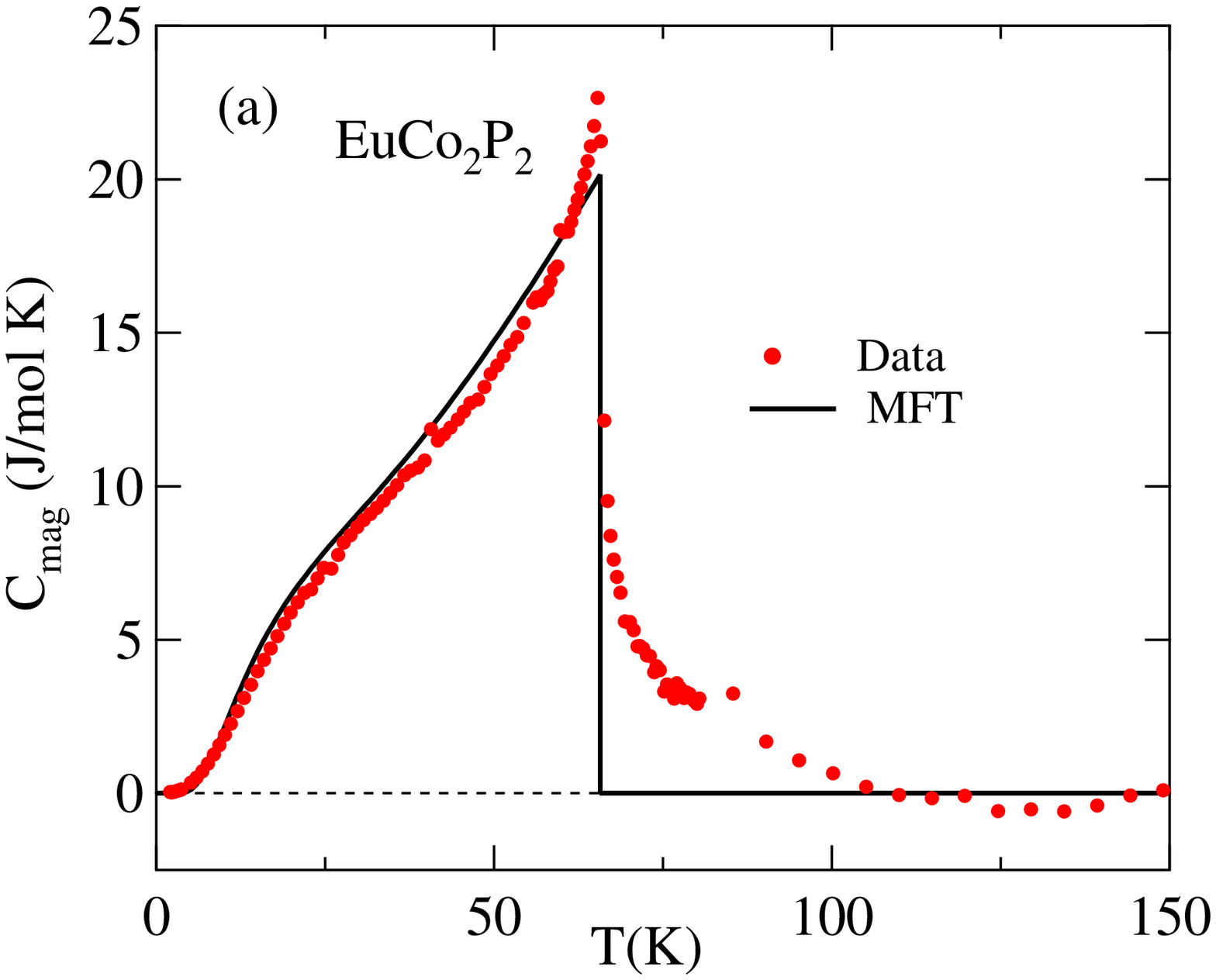}
\includegraphics[width=3.3in]{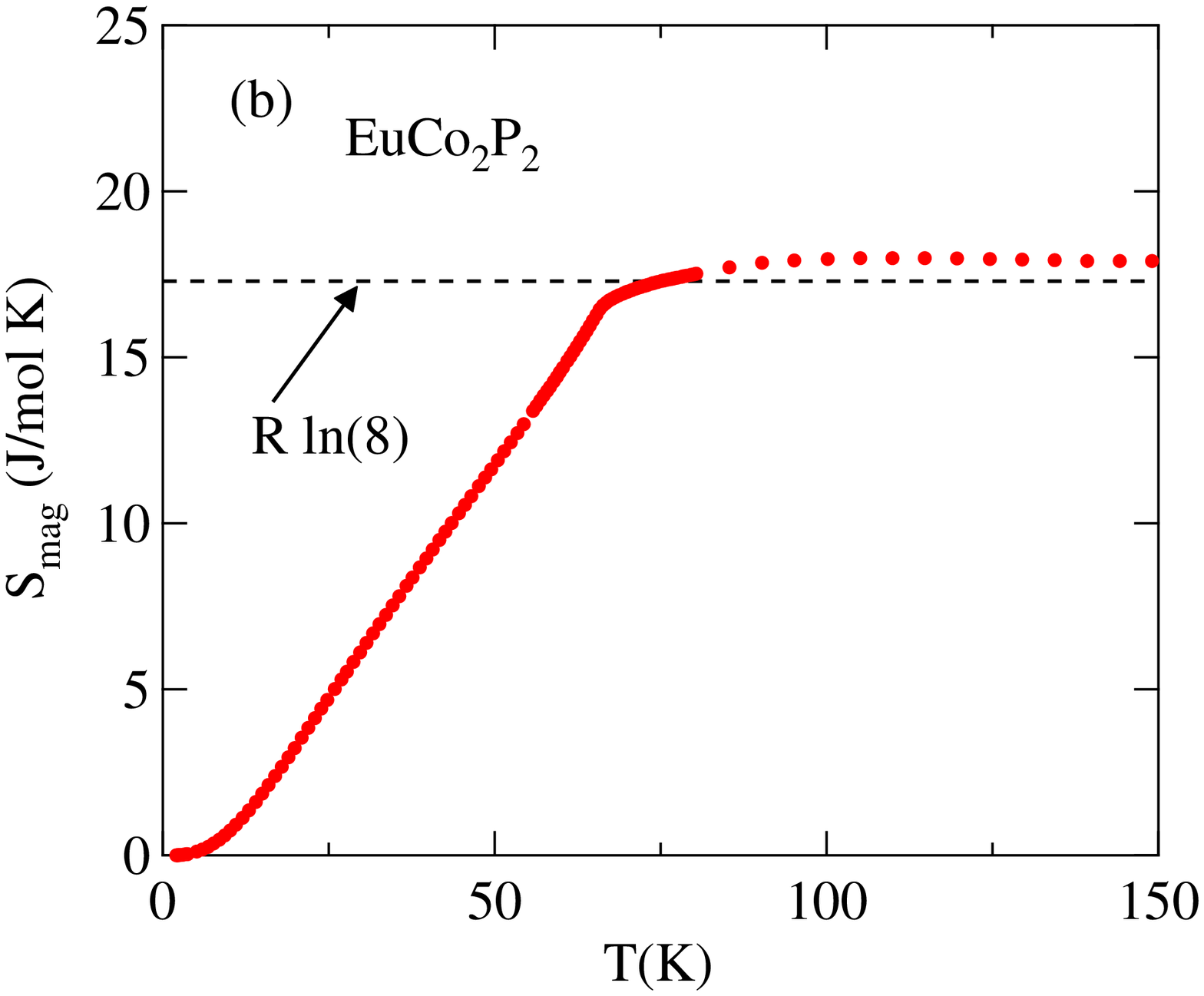}
\caption{(Color online) (a)~Magnetic contribution $C{\rm_{mag}}(T)$ of the Eu spins in \ecp\ obtained after subtraction of the contribution $C{\rm_p}(T)$ of \bcp{}, between 1.8 to 150~K\@. The MFT prediction for $C_{\rm mag}(T)$ with $S=7/2$ and $T_{\rm N} = 65.7$~K is shown as the black solid line.  (b)~Magnetic entropy $S_{\rm mag}(T)$ calculated from $C_{\rm mag}(T)$. The horizontal dashed line is the theoretical high-$T$ limit $S_{\rm mag} = R\ln(2S+1)=17.29$~J/mol~K for $S = 7/2$.}
\label{Fig:EuCo2P2_Cmag}
\end{figure}

\begin{figure}
\includegraphics[width=3.5in]{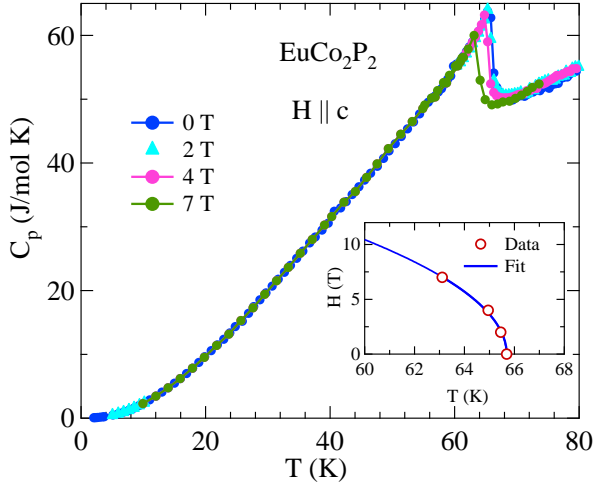}
\caption{(Color online)  Heat capacity $C_{\rm p}$ versus $T$ of \ecp\ at various $H\parallel c$, $C\rm{_p}(H,T)$. Inset: Magnetic $H$-$T$ phase diagram for \ecp{} as determined from $C_{\rm p}(H,T)$ measurements. The solid blue curve is a fit of the four data points by Eq.~(\ref{Eq:HcFit}). }
\label{Fig: EuCo2P2_Cp_field}
\end{figure}

In order to estimate the magnetic contribution $C_{\rm mag}(T)$ of the Eu spins to the $C_{\rm p}(T)$ of \ecp{}, we used the $C_{\rm p}(T)$ data for \bcp\ as a reference.  Because the two compounds have different $\gamma$~values (Table~\ref{Table:Heat_capacity}), the $C_{\rm p}$ data were corrected for the respective electronic $\gamma  T$ terms as shown in Fig.~\ref{Fig: EuCo2P2_Cp}(b).  One sees that the $C_{\rm p}-\gamma  T$ data for the two compounds are now nearly identical above $\approx100$~K\@. To eliminate the residual average deviation of the data between the two compounds in the 100--300~K temperature range we multiplied the $C_{\rm p}-\gamma T$ data for \ecp\ by 1.0046. Then taking the difference between the resulting data for \ecp\ and the $C_{\rm p}-\gamma T$ data for \bcp\ yields the magnetic heat capacity $C_{\rm mag}(T)$ for \ecp\ in Fig.~\ref{Fig:EuCo2P2_Cmag}(a). The nonzero $C_{\rm mag}$ for $T_{\rm N}<T\lesssim100$~K is due to weak dynamic short-range magnetic ordering of the Eu spins above $T_{\rm N}$.

The magnetic entropy $S_{\rm mag}(T)$ is calculated using $S_{\rm mag}(T)=\int_{0}^{T}[C_{\rm mag}(T)/T]dT$ and the result is shown in Fig.~\ref{Fig:EuCo2P2_Cmag}(b). The high-$T$ limit for a mole of spins $S=7/2$ is $R\ln(2S+1)$, as shown by the horizontal dashed line in Fig.~\ref{Fig:EuCo2P2_Cmag}(b). One sees that the high-$T$ $S_{\rm mag}(T)$ data for \ecp\ closely approach this value. The small residual deviation is likely due to a small inaccuracy in the background $C_{\rm p}$ subtraction. The short-range magnetic ordering seen in $C_{\rm mag}$ at $T>T_{{\rm N}}$ in Fig.~\ref{Fig:EuCo2P2_Cmag}(a) represents only a small fraction of the total entropy of the disordered spin system, since the change in $S_{\rm mag}$ from $T_{\rm N}$ to 100~K is found from Fig.~\ref{Fig:EuCo2P2_Cmag}(b) to be only about 7\% of the disordered entropy $R\ln(8)$.

We also measured the $C_{\rm p}(T)$ data for a single crystal of \ecp{} in the $T$~range from 1.8 to 80~K in various magnetic fields applied along the $c$~axis, as shown in Fig.~\ref{Fig: EuCo2P2_Cp_field}. It is seen that the $C{\rm_p}(H,T)$ shows a decrease in $T_{\rm N}$ by only $\approx 3$~K upon varying $H$ from 0 to 7~T\@. The inset of Fig.~\ref{Fig: EuCo2P2_Cp_field} shows the $H$-$T$ phase diagram determined from the $H$ dependence of $T{\rm _{N}}$ and the solid blue curve is a fit of the data by the empirical function
\bse
\label{Eqs:HcSrCo2P2}
\be
H_{\rm c}(T) = H_0 \left(1-\frac{T}{T{\rm_N}}\right)^{1/2} \quad (T\to T_{\rm N}),
\label{Eq:HcFit}
\ee
where  $T_{\rm N}$ is fixed to the above value 65.7~K\@.  The fitting parameter is
\be
H_0 = 36~{\rm T}.
\label{Eq:HcPars}
\ee
\ese
From Eq.~(\ref{Eq:HcFit}) one can invert the axes to obtain $T_{\rm N}(H)$ as
\be
T_{\rm N}(H) = T_{\rm N}(0)\left[1-\left(\frac{H}{H_0}\right)^2\right]\quad (T\to T_{\rm N}).
\ee

According to MFT \cite{Johnston2015}, for fields applied along the $c$ axis, which is perpendicular to the plane of the ordered moments in $H=0$, the AFM-PM phase boundary $H_{\rm c}(t)$ is given by
\bse
\label{Eqs:Hc}
\be
H_{\rm c}(t) = H_{\rm c}(0)\bar{\mu}_0(t),
\label{Eq:Hc(T)MFT}
\ee
where
\bea
\bar{\mu}_0(t) &=& \frac{\mu_0(t)}{\mu_{\rm sat}} = \frac{\mu_0(T)}{gS\mu_{\rm B}},\qquad t = \frac{T}{T_{\rm N}},
\eea
$\bar{\mu}_0(T)$ is the reduced AFM ordered moment in $H=0$ which equals unity at $T=0$, $\mu_{\rm sat}= gS\mu_{\rm B}$ is the saturation moment of the spin~$S$ and $t$ is the reduced temperature.  For $t\to1$, one obtains \cite{Johnston2015}
\be
\bar{\mu}_0(t) = \frac{\sqrt{10/3}\ (1+S)}{\sqrt{1+2S+2S^2}}\ (1-t)^{1/2} \qquad (t\to1).
\ee
Then Eq.~(\ref{Eq:Hc(T)MFT}) becomes
\be
H_{\rm c}(t) = H_{\rm c}(0)\frac{\sqrt{10/3}\ (1+S)}{\sqrt{1+2S+2S^2}}\ (1-t)^{1/2} \qquad (t\to1).
\label{Eq:Hc(T)MFT2}
\ee
Thus the temperature dependence of the experimental data described by Eq.~(\ref{Eq:HcFit}) agrees with the MFT prediction in Eq.~(\ref{Eq:Hc(T)MFT2}).  

For $S=7/2$, Eq.~(\ref{Eq:Hc(T)MFT2}) gives
\be
H_{\rm c}(t) = 1.441 H_{\rm c}(0) \qquad (S=7/2,\ t\to1).
\label{Eq:HcS72}
\ee
\ese
Comparing Eqs.~(\ref{Eqs:HcSrCo2P2}) and (\ref{Eq:HcS72}) gives
\be
H_{\rm c}(0) \approx \frac{36~{\rm T}}{1.441} = 25~{\rm T} \quad {\rm for\ EuCo_2P_2}.
\label{Eq:Hc(0)value}
\ee
This value is reasonably close to the value of $\approx 28$~T obtained from a linear extrapolation of $M_c(H,T=2~{\rm K})$ in Fig.~\ref{Fig:EuCo2P2_MH} below from the maximum experimental field of 14~T to the field $H_{\rm c}$ at which the saturation magnetization is attained.

\section{\label{Sec:MChi} Magnetization and Magnetic Susceptibility}

\subsection{Magnetic Susceptibility}

\begin{figure}
\includegraphics[width=3.5in]{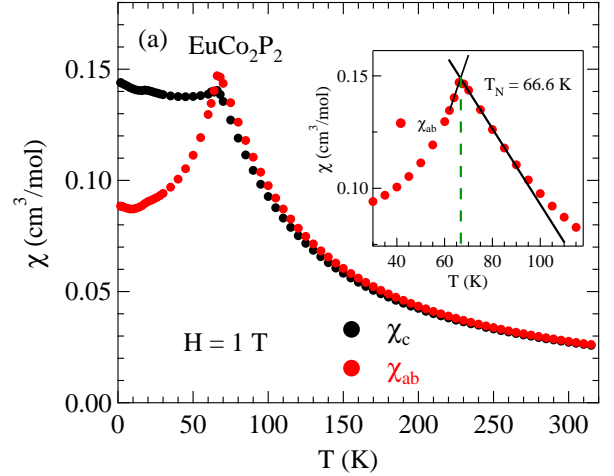}
\includegraphics[width=3.5in]{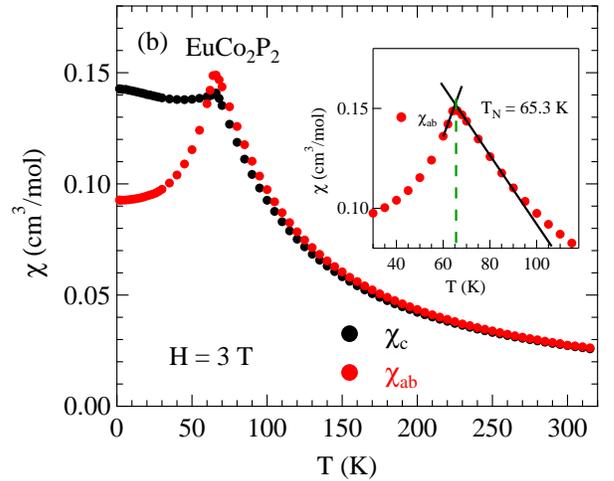}
\caption{(Color online) Zero-field-cooled (ZFC) magnetic susceptibility $\chi\equiv M/H$ of a \ecp{} single crystal in the temperature $T$ range between 1.8 and~320~K measured with magnetic fields $\mathit{H}$ applied along the $\mathit{c}$ axis ($\chi_{c},\ H\parallel c$) and in the $ab$ plane ($\chi_{ab},\ H\parallel ab$) for (a)~$H=1$~T and (b) $H=3$~T\@. The insets in~(a) and~(b) show $\chi_{ab}(T)$ around $T_{\rm N} =  66.6(2)$~K and 65.3(1)~K, respectively.}
\label{Fig:EuCo2P2_chi}
\end{figure}

The zero-field-cooled (ZFC) $\chi\equiv M/H$ of a \ecp{} single crystal as a function of~$T$ measured at $H = 1$~T and 3~T applied along the $c$ axis ($\chi_{c},\, H\parallel c$) and in the $ab$ plane ($\chi_{ab},\, H\parallel ab$) are shown in Figs.~\ref{Fig:EuCo2P2_chi}(a) and \ref{Fig:EuCo2P2_chi}(b), respectively.  As shown in the inset of Fig.~\ref{Fig:EuCo2P2_chi}(a), a sharp cusp in $\chi_{ab}$ is observed in $H=1$~T at $T=66.6(2)$~K that we identify as $T{\rm{_N}}$, a value in good agreement with the previous reports \cite{Morsen1988, Nakama2010, Reehuis1992}.  The $\chi_{ab}(T)$ data for $H=3$~T in the inset of Fig.~\ref{Fig:EuCo2P2_chi}(b) shows a reduction of $T_{\rm N}$ by about 1.3~K, consistent with that obtained from $C_{\rm p}(T)$ measurements in the inset of  Fig.~\ref{Fig: EuCo2P2_Cp_field}.  The $\chi_{c}(T)$ data in Fig.~\ref{Fig:EuCo2P2_chi} shows an abrupt change in slope at $T{\rm{_N}}$ and becomes nearly $T$-independent at lower~$T$\@.  Further, one observes from Fig.~\ref{Fig:EuCo2P2_chi} that $\chi_{ab}/\chi_{c}\approx $1.05 at $T = T_{\rm{N}}$ which arises from anisotropy in the system (see below). Similar features have been reported in other collinear and noncollinear AFM compounds \cite{Anand2015, Anand2014}. 

The anisotropy in the $\chi(T<T\rm_{N})$ data in Fig.~\ref{Fig:EuCo2P2_chi} indicates that the $ab$~plane is an easy plane. In addition, the nonzero limit of $\chi_{ab}(T\to0)$ indicates that \ecp{} is either a collinear AFM with orthogonal domains or a planar noncollinear AFM structure \cite{Johnston2012, Johnston2015, Anand2015}.  The previous neutron diffraction study on \ecp{} \cite{Reehuis1992} showed that the latter possibility is the correct one, namely an incommensurate AFM helical structure in which Eu spins are aligned ferromagnetically within the $ab$~plane and where the helix axis is the $c$~axis.  

\begin{figure}
\includegraphics[width=3.5in]{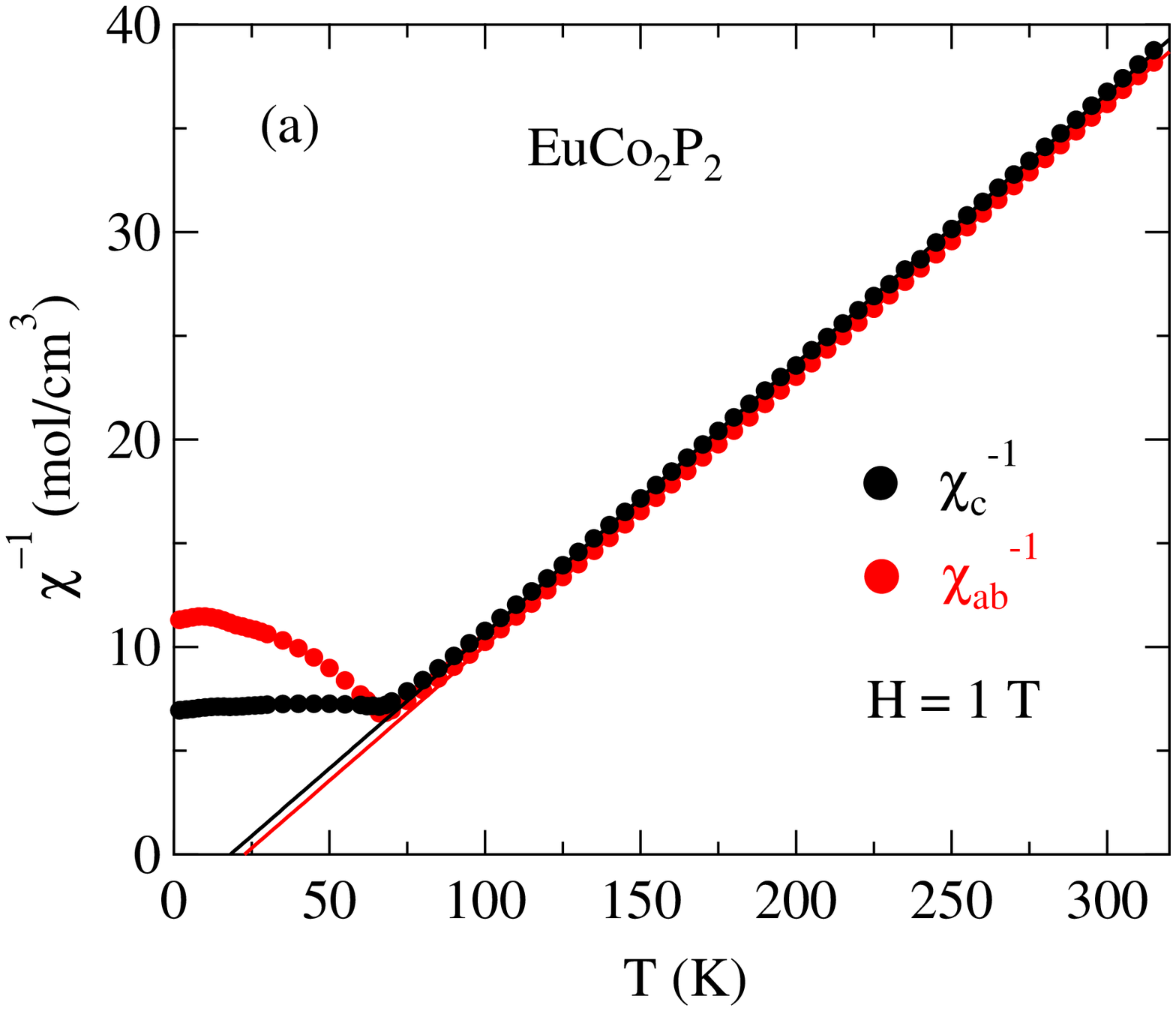}
\includegraphics[width=3.5in]{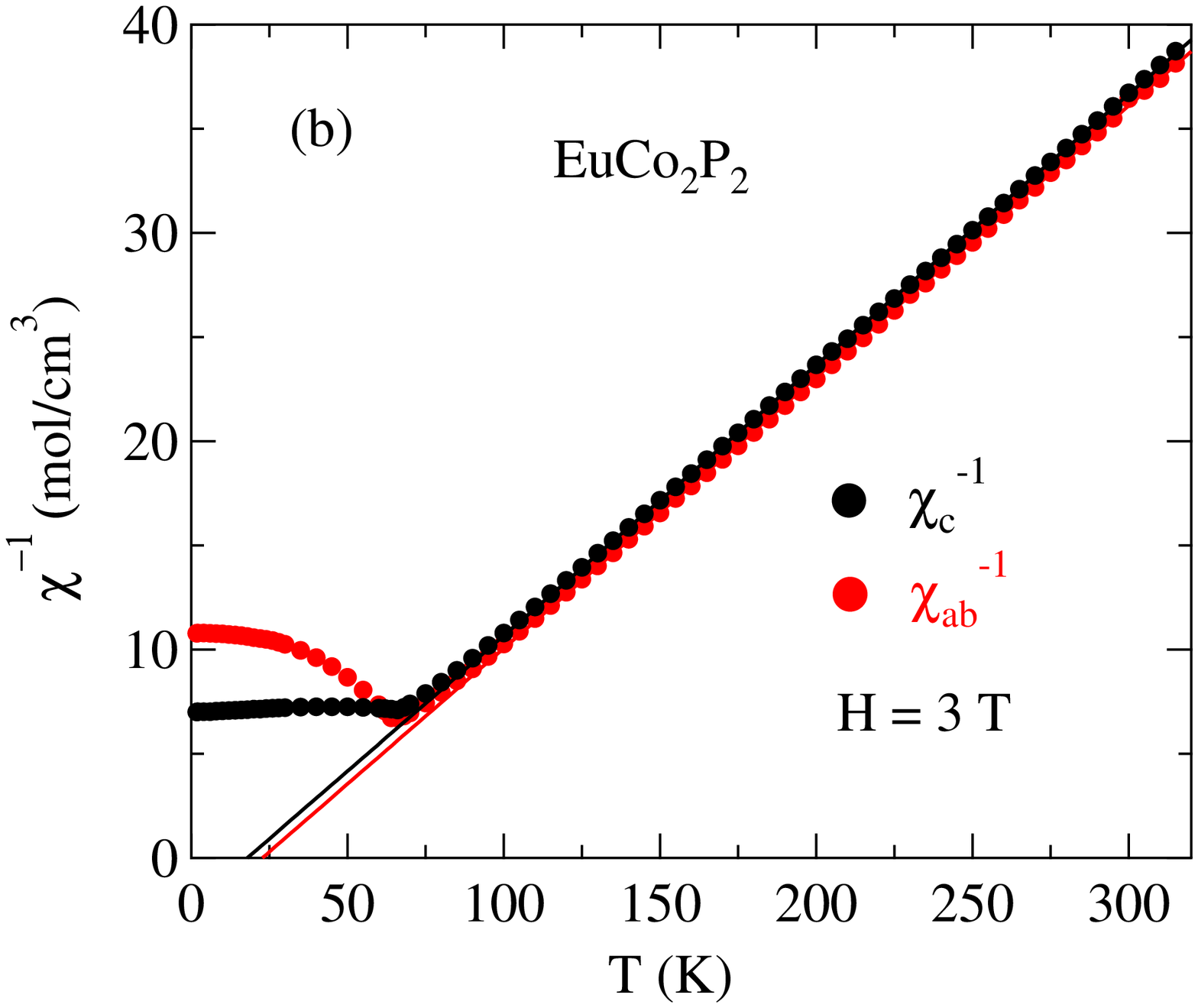}
\caption{(Color online) Zero-field-cooled (ZFC) inverse magnetic susceptibility $\chi^{-1}$ for single-crystal \ecp\ in the temperature $T$ range from~1.8 to 320~K measured with magnetic fields (a)~$H=1$~T and (b)~$H=3$~T applied along the $c$~axis ($\chi_{c}^{-1},\, H\,\parallel c$) and in the $ab$~plane ($\chi_{ab}^{-1},\, H\,\parallel ab$). The straight lines are respective fits of the $\chi^{-1}(T)$ data by the Curie-Weiss law.}
\label{Fig:EuCo2P2_InvChi}
\end{figure}

The $\chi^{-1}(T)$ of \ecp {} measured in $H$=1~T and 3~T applied along the $c$ axis ($\chi_{c}^{-1},\, H\,\parallel c$) and in the $ab$ plane ($\chi_{ab}^{-1},\, H\,\parallel ab$) are shown in Figs.~\ref{Fig:EuCo2P2_InvChi}(a) and~\ref{Fig:EuCo2P2_InvChi}(b), respectively. The high-temperature (100~$<{T}<$~320~K) magnetization data in the PM state are fitted by the Curie-Weiss law~(\ref{Eq:CW-law}), where the Curie constant is given by \cite{Kittel2005}
\begin{equation}
C= \frac{Ng^2S(S+1)\mu_{\rm B}^2}{3k_{\rm B}},
\label{Eq:Curie constant}
\end{equation}
where $N$ is the number of spins. The Curie-Weiss fits are shown as the straight lines in Figs.~\ref{Fig:EuCo2P2_InvChi}(a) and~\ref{Fig:EuCo2P2_InvChi}(b). The fitted parameters $C$ and $\theta_{\rm p}$ together with $T_{\rm N}$ and $f \equiv\theta_{\rm p}/T_{\rm N}$ are listed in Table~\ref{Table: curieFit}. The values of $C$ are same for both field directions and are within 1\% of the theoretical value $7.88~{\rm cm^3\,K/mol}$ for Eu$^{+2}$ spins with $S = 7/2$ and $g=2$. This indicates negligible contribution from the Co as also indicated from the previous neutron diffraction study \cite{Reehuis1992}.

The difference 
\be
\theta_{ab}-\theta_{c} = [23.0(3) - 18.2(3)]~\rm{K} = 4.8(4)~\rm{K}
\label{Eq:theta_diff}
\ee
arises mainly from magnetic dipole interactions between the Eu spins as discussed in Sec.~\ref{Sec:MDIs}. The spherical average of the fitted Weiss temperatures is $\theta_{\rm p,ave}=21.4(3)$~K, in agreement with the previous result of 20(2)~K (Ref.~\onlinecite{Morsen1988}) for a polycrystalline sample and its positive value indicates predominantly FM exchange interactions between the Eu spins \cite{Johnston2015}.

\begin{table}
\caption{\label{Table: curieFit} AFM ordering temperature $T_{\rm N}$ and parameters obtained for \ecp\ single crystals obtained from fits from 100 to 320~K of the $\chi^{-1}(T)$ data in Fig.~\ref{Fig:EuCo2P2_InvChi} by the Curie-Weiss law.  The parameter $f$ is $f\equiv\theta_{\rm p}/T_{\rm N}$. }
\begin{ruledtabular}
\begin{tabular}{lcccc}
Field and       		& $T_{\rm N}$	& 		$C$        	& $\theta_{\rm p}$  &  $f~(T_{\rm N} = 66.5$~K)	\\
direction      	& (K)  		& (${\rm cm^3\,K/mol}$)	&         (K)  			\\
\hline 
$H = 1$~T									\\
$H\parallel ab$  	& 66.6(2) 	& 7.67(1)  			& 23.0(3)  			& 0.346  	\\
$H\parallel c$  	& 66.6(1)  		& 7.69(1)  			& 18.2(3)  			& 0.274   \\
				&			&					& 21.4(3)\footnotemark 			& 0.320\\
\hline 
$H = 3$~T									\\
$H\parallel ab$  	& 65.3(1)  		& 7.67(2)  			& 22.8(3)  		& 0.349	\\
$H\parallel c$  	& 	  		& 7.69(1)  			& 18(2)  			& 0.276	\\
\end{tabular}
\end{ruledtabular}
\footnotetext{spherical average}
\end{table}

\subsection{Magnetization versus Magnetic Field Isotherms}

\begin{figure}
\includegraphics[width=3.5in]{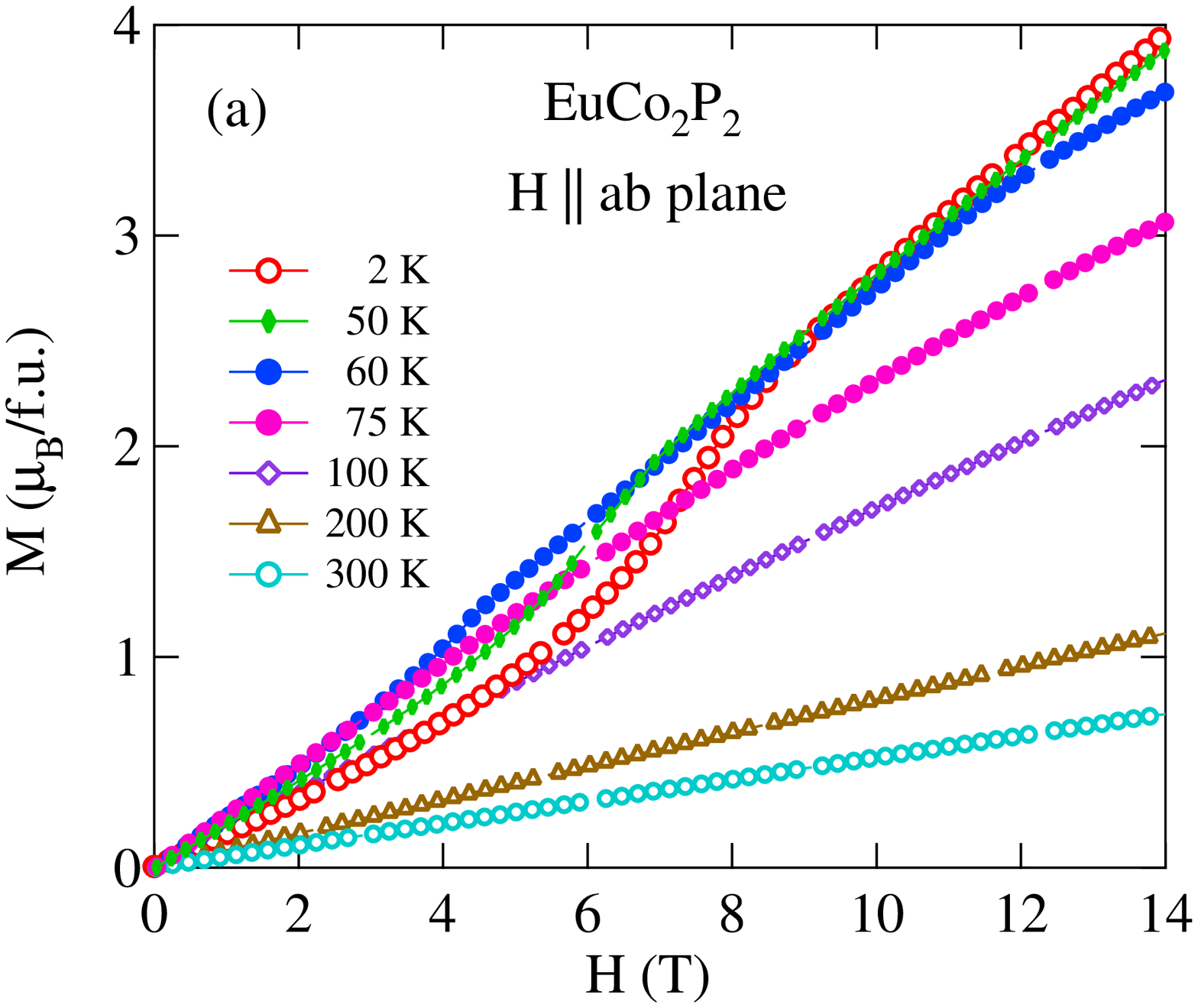}
\includegraphics[width=3.5in]{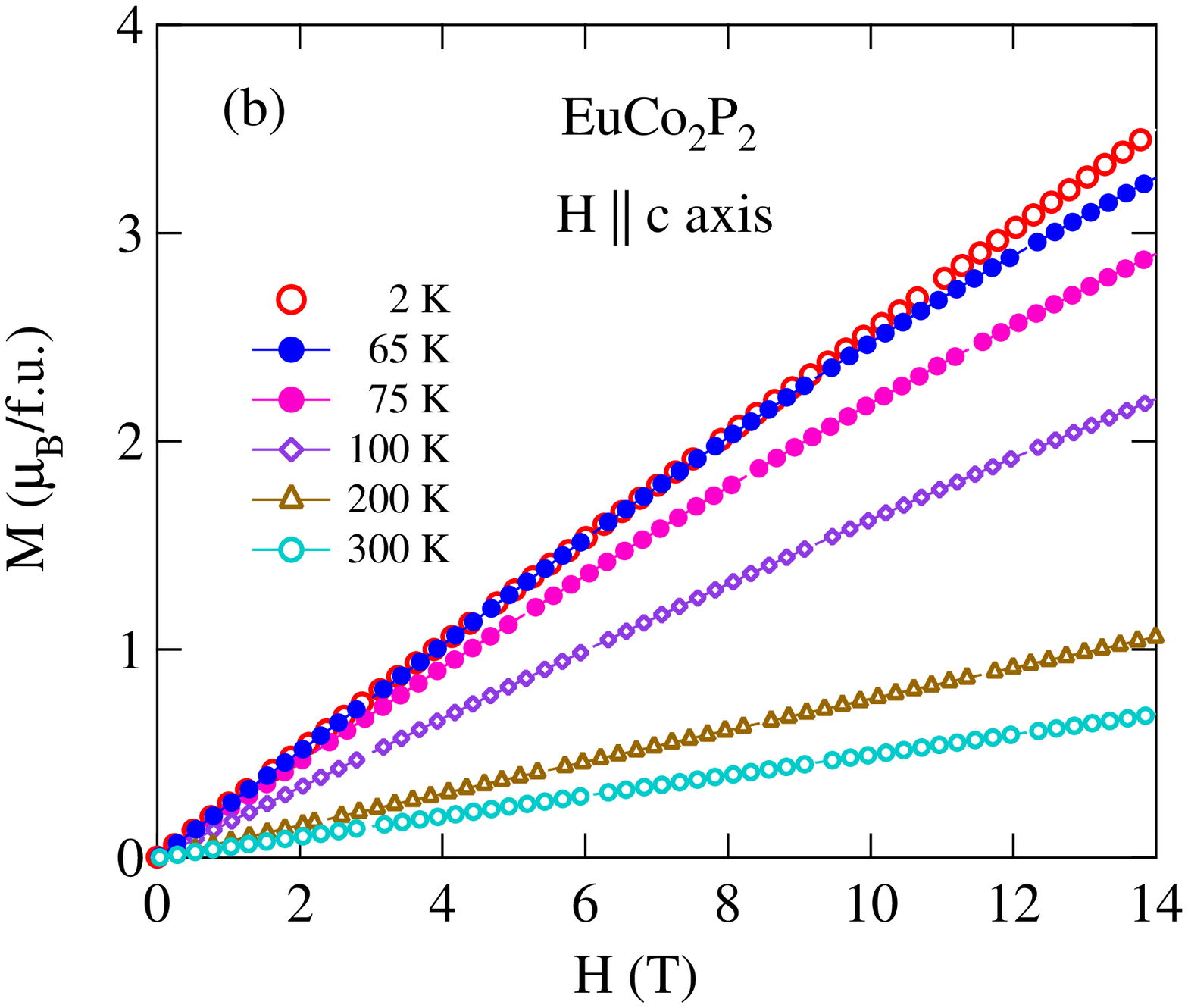}
\caption{(Color online) Magnetization $M$ versus applied magnetic field $H$ isotherms for an \ecp\ crystal at the indicated temperatures~$T$ for (a) $H \parallel ab$ and (b) $H \parallel c$. The isotherms in~(a) for $T<T_{\rm N}$ exhibit metamagnetic transitions.}
\label{Fig:EuCo2P2_MH}
\end{figure}

\begin{figure}
\includegraphics[width=3.5in]{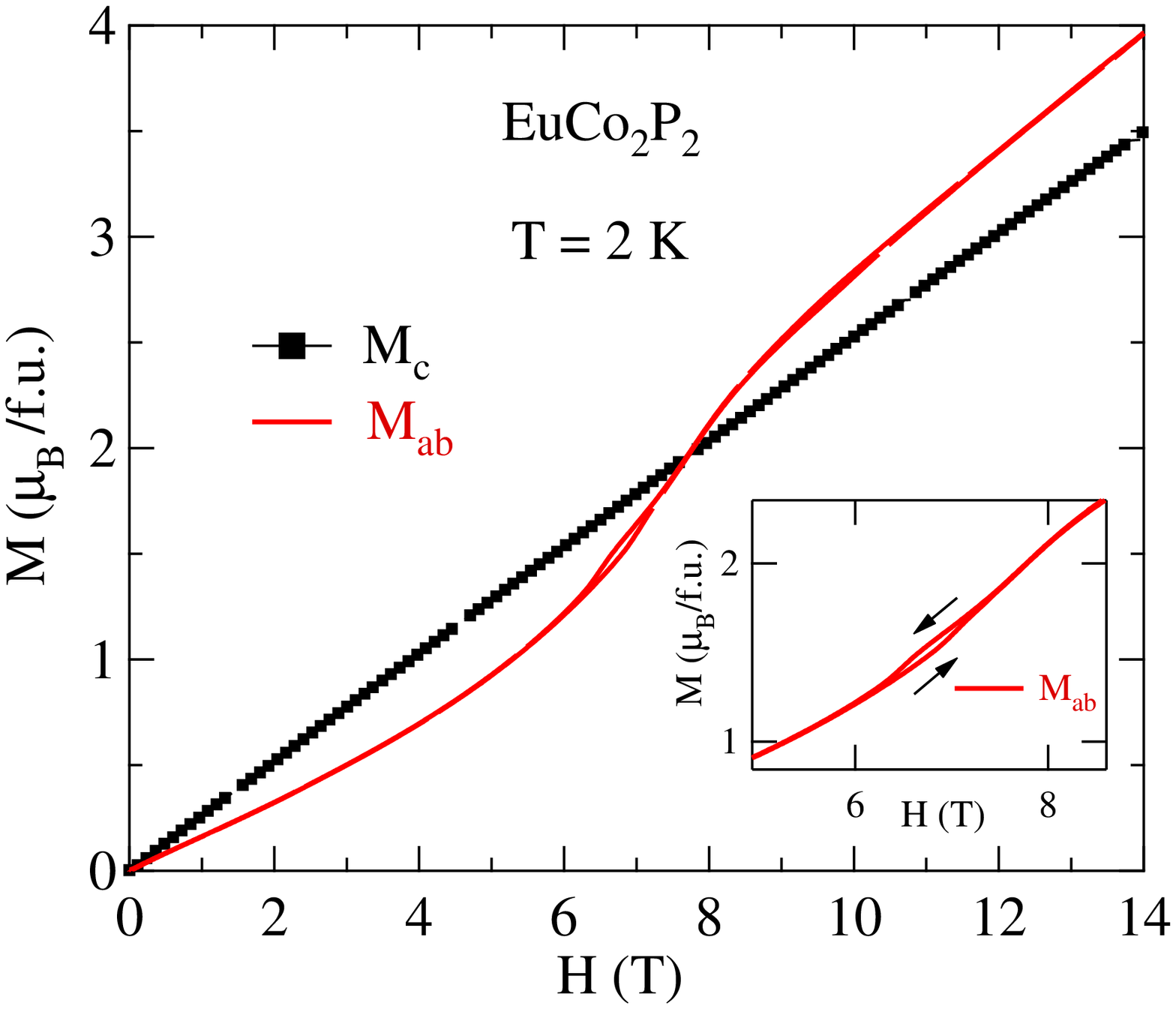}
\caption{(Color online) Isothermal magnetization $M$ of \ecp\ versus magnetic field $H$ at $T = 2$~K for $H \parallel ab$  and $H \parallel c$.  Inset: expanded view of hysteresis in the $M_{ab}(H)$ data.}
\label{Fig:EuCo2P2_MH 2K}
\end{figure}

Isothermal $M(H)$ data for \ecp\ measured at temperatures from 2 to 300~K with $H$ applied in the $ab$ plane ($M_{ab},H\parallel ab$) and along the $c$ axis ($M_{c},H\parallel c$) are shown in Figs.~\ref{Fig:EuCo2P2_MH}(a) and~\ref{Fig:EuCo2P2_MH}(b), respectively. For clarity, the data at $T=2$~K are shown separately in Fig.~\ref{Fig:EuCo2P2_MH 2K}, where the $M_{c}(H)$ data are nearly linear in field as predicted at $T\ll T_{\rm N}$ by MFT for a helix with the applied field along the helix axis \cite{Johnston2015}. s A linear extrapolation of the $M_{c}(H)$ data to the saturation magnetization $M_{\rm sat} = 7~\mu_{\rm B}$/Eu yields the critical field $H_c(T\to0)\approx 28$~T\@.  As $T$ increases above~$T_{\rm N}$, the $M_{c}(H)$ data in Fig.~\ref{Fig:EuCo2P2_MH}(b) show a small negative curvature at $T=60$~K and 65~K\@. At even higher temperatures, a proportional behavior of $M_{c}(H)$ is eventually observed.

\begin{figure}
\includegraphics[width=3.5in]{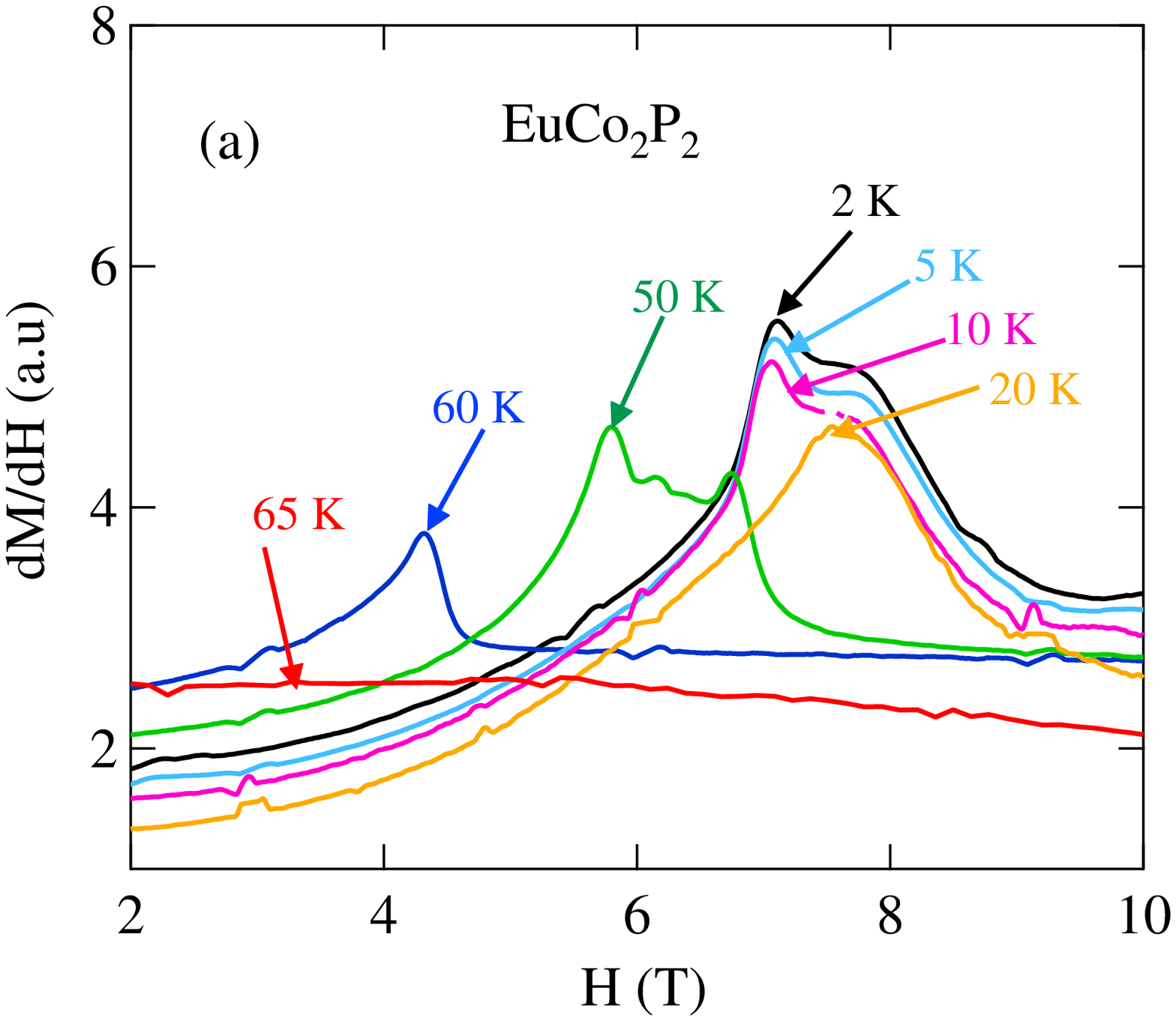}
\includegraphics[width=3.5in]{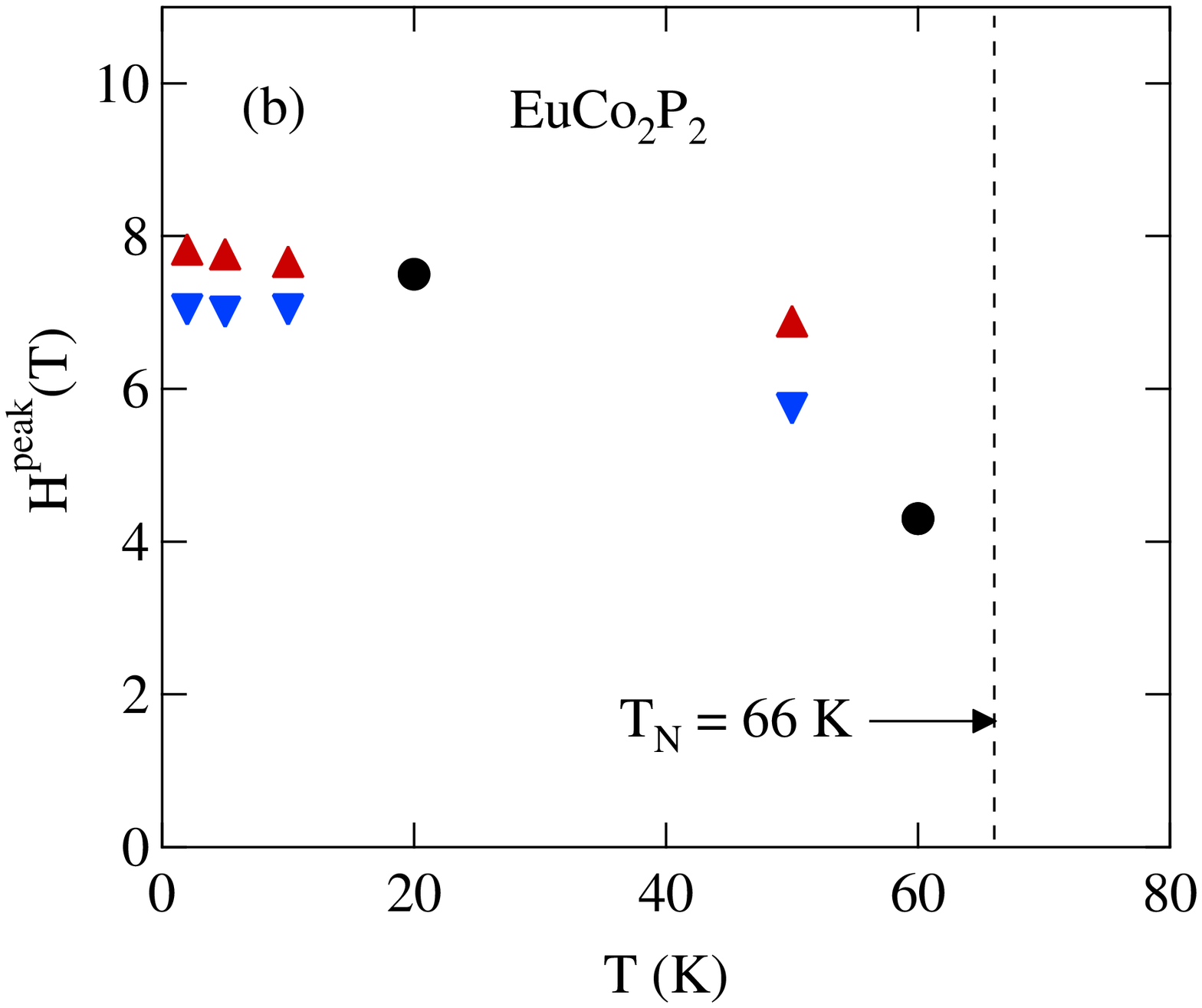} 
\caption{(Color online) (a)~Derivative $dM/dH$ versus~$H$ for several temperatures~$T$ as indicated.  (b)~Magnetic peak fields $H^{\rm peak}$ of the $dM/dH$ data in~(a) versus $T$\@.  Four of the plots in~(a) have two peak temperatures associated with them at a higher (upwards pointing filled triangle) and lower (downwards pointing filled triangle) fields.  The other two curves only show a single peak at each temperature (filled circles).}
\label{Fig:dM_vs_dH}
\end{figure}

On the other hand, the $M_{ab}(H)$ isotherms in Fig.~\ref{Fig:EuCo2P2_MH}(a) for $T\leq60$~K show a metamagnetic transition at each temperature, where the transition field decreases from a maximum value of $\approx7$~T at 2~K to~0 at $T_{\rm N}$\@. A small hysteresis in $M_{ab}(H)$ at 2~K is observed upon field cycling as shown in the inset of Fig.~\ref{Fig:EuCo2P2_MH 2K}, suggesting a first-order transition. To more clearly study these transitions, $dM_{ab}/dH$ versus $H$ isotherms obtained from the three 2~K to 60~K isotherms in Fig.~\ref{Fig:EuCo2P2_MH}(a) and also from additional isotherms not shown there are plotted in Fig.~\ref{Fig:dM_vs_dH}(a).  The data for $T\leq 60$~K exhibit distinct peaks at fields $H^{\rm peak}$ that decrease with increasing~$T$ and disappear at~$T_{\rm N}$ as shown in Fig.~\ref{Fig:dM_vs_dH}(b).  Four of the six peaks show two distinct features at closely-spaced fields.  The data in these figures suggest multiple changes in the AFM stucture with increasing field.

\subsection{\label{Sec:MDIs} Influence of Magnetic Dipole Interactions and Single-Ion Anisotropy on the Weiss and N\'eel Temperatures}

Here we estimate the influence of magnetic dipole interactions (MDIs) on the anisotropic Weiss temperatures $\theta_{{\rm p}ab}$ and $\theta_{{\rm p}c}$ and their influence on the N\'eel temperature $T_{\rm N}$.  The crystal for which the above magnetization measurements were obtained was a nearly square flat plate with dimensions $(1.69^2\times0.46)$~mm$^3$ for the $ab$~plane and $c$~axis directions, respectively.  From these dimensions one obtains the magnetometric demagnetizing factors \cite{Aharoni1998}
\be
N_{{\rm d}\,ab} = 0.183, \qquad N_{{\rm d}\,c} = 0.634,
\label{Eq:Nd}
\ee
which are defined here as in the SI system of units where $0 \leq N_{{\rm d}\,\alpha} \leq 1$, $\sum_{\alpha=1}^3 N_{{\rm d}\,\alpha} = 1$, and $\alpha \equiv ab$ or~$c$.  

\subsubsection{Weiss Temperatures}

The contribution $\theta_{{\rm MDI}\alpha}$ of the MDIs to the Weiss temperature measured in the $\alpha$ principal-axis direction is \cite{Johnston2016}
\be
\theta_{{\rm MDI}\alpha} = \frac{C_1}{a^3}\left[\lambda_{{\bf 0}\alpha} + \frac{4\pi}{V_{\rm spin}/a^3}\left(\frac{1}{3}-N_{{\rm d}\,\alpha}\right)\right],
\label{Eq:thetaMDI}
\ee
where $\lambda_{{\bf 0}\alpha}$ is the eigenvalue of the MDI tensor for the PM state with $H\parallel\alpha$ for the known $c/a$ ratio of the bct Eu sublattice, $V_{\rm spin}$ is the volume per spin and $C_1$ is the single-spin Curie constant 
\be
C_{\rm 1} = \frac{g^2S(S+1)\mu_{\rm B}^2}{3k_{\rm B}}.
\ee
The values of $\lambda_{{\bf 0}\alpha}$ for $c/a=3.014$ and $ka=0.880\pi$\,rad for \ecp\ calculated by direct summation for neighboring moments within a radius of $100a$ of the central moment (2,779,450 neighboring spins) are
\be
\lambda_{{\bf 0}ab} = 4.5058, \qquad \lambda_{{\bf 0}c} = -2\lambda_{{\bf 0}ab}.
\label{Eq:lambdaVals}
\ee
The most positive eigenvalue corresponds to the lowest-energy moment direction, which here lies within the $ab$~plane, consistent with the helix AFM structure of \ecp\ below $T_{\rm N}$\@.  Then using $V_{\rm spin} = a^2c/2$, $g=2$ and the lattice parameters in Table~\ref{Table: curieFit} gives
\bea
\theta_{{\rm MDI}ab} &=& 1.41\,{\rm K},\qquad \theta_{{\rm MDI}c} = -2.82\,{\rm K},\nonumber\\*
&&  \theta_{{\rm MDI}ab} - \theta_{{\rm MDI}c} = 4.23\,{\rm K}.
\label{Eq:thetaMDDiff}
\eea
The value of $\theta_{{\rm MDI}ab} - \theta_{{\rm MDI}c}$ is slightly smaller than the measured value of 4.8(4)~K in Eq.~(\ref{Eq:theta_diff}).

Another source of anisotropy is single-ion (SI) anisotropy of the Eu$^{+2}$ spins-7/2 with a contribution to the Hamiltonian written as $-DS_z^2$, where the $z$ axis is the uniaxial $c$ axis here and $D$ is the anisotropy constant which is negative for planar anisotropy.  The anisotropy and Zeeman parts of the Hamiltonian are
\be
{\cal H} = -DS_z^2 + g\mu_{\rm B} {\bf S}\cdot{\bf H}
\ee
for ${\bf H} = H_x\hat{\bf i}$ or ${\bf H} = H_z\hat{\bf k}$.  Diagonalizing the Hamiltonian matrix for $S= 7/2$ for the two field directions and calculating the anisotropic $\chi$ at high~$T$ therefrom, we find the contributions to the anisotropic Weiss temperatures arising from the SI anisotropy to be
\bea
\theta_{{\rm SI}ab} &=& -\frac{2D}{k_{\rm B}},\quad \theta_{{\rm SI}c} = -2\theta_{{\rm SI}ab},\nonumber\\*
&& \theta_{{\rm SI}ab} - \theta_{{\rm SI}c} = -\frac{6D}{k_{\rm B}}.
\eea
In order to explain the small difference $\sim 0.6$~K between the $\theta_{ab} - \theta_{c}$ values in Eqs.~(\ref{Eq:theta_diff}) and~(\ref{Eq:thetaMDDiff}) requires $D/k_{\rm B} \sim -0.1$~K, i.e., a planar anisotropy.  This $D$ value seems reasonable.  For example, using ESR measurements similar magnitudes $|D|/k_{\rm B} \sim 0.1$--0.3~K have been inferred for Gd$^{+3}$ with $S=7/2$ in ${\rm Gd_2Ti_2O_7}$ (Ref.~\onlinecite{Glazkov2005}) and ${\rm Gd_2Sn_2O_7}$  (Ref.~\onlinecite{Glazkov2006}) for Gd$^{+3}$ with $S=7/2$.

\subsubsection{N\'eel Temperature}

The contribution $T_{\rm N\,MDI}$ from MDIs to the N\'eel temperature for AFM ordering is \cite{Johnston2016}
\be
T_{\rm N\,MDI} = \frac{C_1\lambda_{\bf k\alpha}}{a^3},
\label{Eq:TNMDI}
\ee
where {\bf k} is the AFM wave vector and $\alpha=ab$ is the ordering plane for a helix.  Taking ${\bf k} = 0.855 (2\pi/c)\hat{\bf c}$ for the helix wave vector one obtains
\be
\lambda_{{\bf k}\,ab} = 4.51.
\ee
Using Eq.~(\ref{Eq:TNMDI}) then gives
\be
T_{\rm N\,MDI} = 1.11~{\rm K}.
\ee

The contribution $T_{{\rm NSI}ab}$ of the single-ion anisotropy to the measured~$T_{\rm N}$ for half-integer spins and AFM ordering in the $ab$~plane with $D/(k_{\rm B}T_{\rm N}) \ll 1$ is obtained as
\be
T_{{\rm NSI}ab} = -\frac{DS(S+1)}{3k_{\rm B}},
\ee
where $D<0$ for in-plane ordering as discussed above.  Using $S=7/2$ and $D/k_{\rm B} = -0.1$~K from the previous section gives
\be
T_{{\rm NSI}ab} = 0.5~{\rm K},
\ee
which is smaller than the contribution from MDIs.  Thus the total enhancement of $T_{\rm N}$ due to both the MDIs and single-ion anisotropy is about 1.6~K, or about 2\% of $T_{\rm N} = 66$~K\@.

\subsection{Fit of $\chi_{ab}(T\leq T_{\rm N})$ by Molecular Field Theory}

In order to fit the $ab$-plane susceptibility by the MFT for Heisenberg AFMs in Refs.~\onlinecite{Johnston2012} and~\onlinecite{Johnston2015}, it is convenient to first remove the influences of the various anisotropies (shape, MDI and SI anisotropies) on $\chi$. In general, the contribution $\chi_{J\alpha}$ of Heisenberg interactions to the $\chi_\alpha(T)$ $(\alpha = ab$ or~$c$) in the presence of these anisotropies is given by
\bse
\label{Eqs:ChiJChialpha}
\be
\frac{1}{\chi_\alpha(T)} = \frac{1}{\chi_{J\alpha}(T)} + A_\alpha,
\ee
yielding
\be
\chi_{J\alpha}(T) = \frac{\chi_\alpha(T)}{1 - A_\alpha\chi_\alpha(T)}.
\label{Eq:ChiJalpha}
\ee
where $A_\alpha$ is a $T$-independent constant for each $\alpha$.  Since the net magnetic anisotropy tensor is traceless, one has
\be
2A_{ab} + A_c = 0, \quad{\rm or}\quad A_{ab}=-A_c/2.
\label{Eq:lambdaSum}
\ee
\ese
From Eqs.~(\ref{Eqs:ChiJChialpha}) one obtains
\bea
\frac{1}{\chi_c} - \frac{1}{\chi_{ab}} &=& \frac{1}{\chi_{Jc}(T)} - \frac{1}{\chi_{Jab}(T)} + A_c -A_{ab}\nonumber \\
 &=& \frac{1}{\chi_{Jc}(T)} - \frac{1}{\chi_{Jab}(T)} + \frac{3}{2}A_c.\label{Eq:Delta1chi}
\eea

\begin{figure}
\includegraphics[width=3.4in]{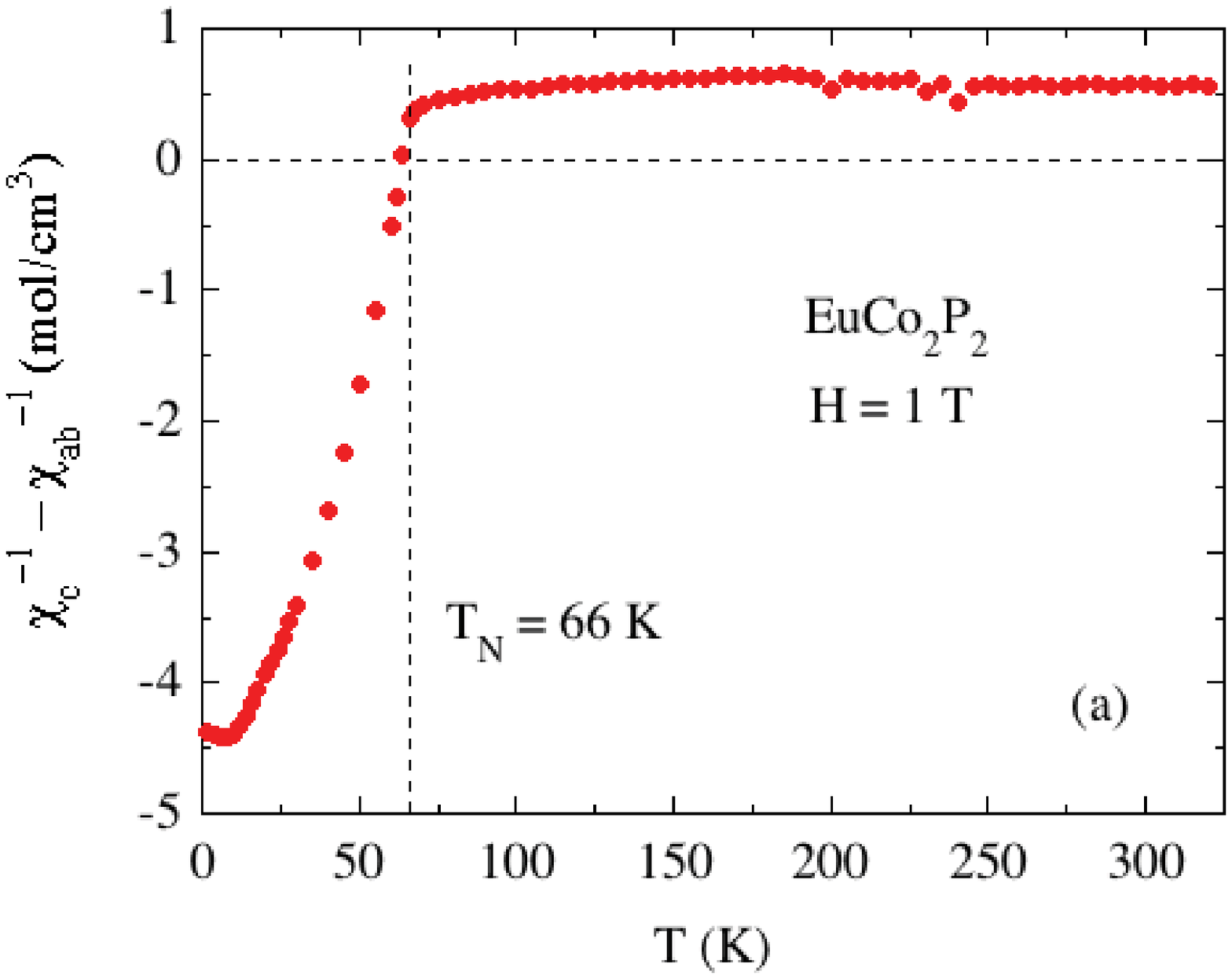}
\includegraphics[width=3.4in]{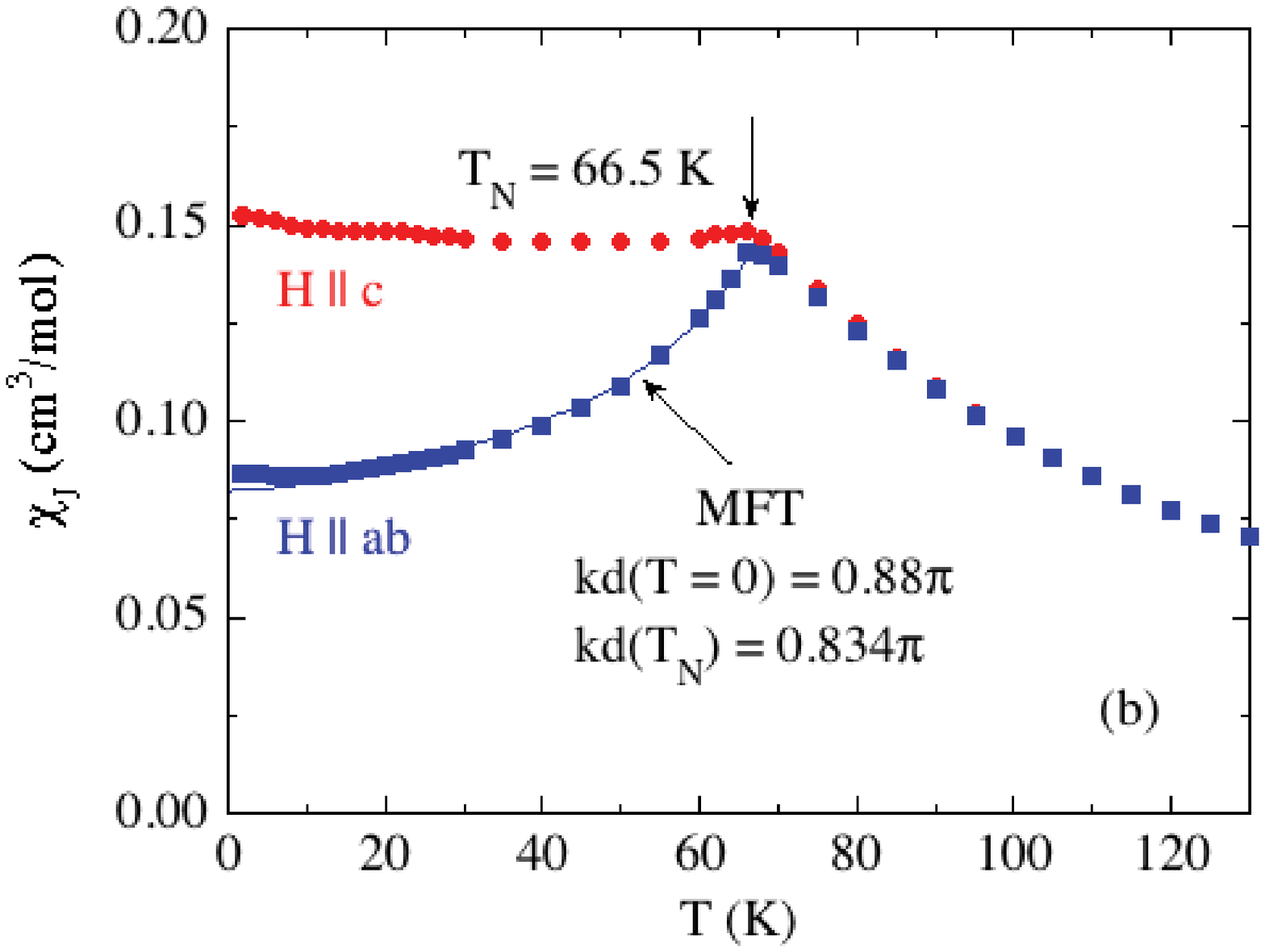}
\caption {(Color online) (a)~$\chi_c^{-1}-\chi_{ab}^{-1}$ versus temperature~$T$ obtained from the data for $H=1$~T in Fig.~\ref{Fig:EuCo2P2_InvChi}(a).  (b)~$\chi_J(T)$ versus $T$ for $H\parallel ab$ and $H\parallel c$ in $H=1$~T after correction for the anisotropies.  The fit of $\chi_{Jab}(T)$ for $T\leq T_{\rm N}$ by the MFT prediction for a helix in Eqs.~(\ref{Eqs:Chixy}) is shown as the blue curve.}
\label{Fig:Chi_1T_EuCo2P2_TN66_ka88}
\end{figure}

In the Curie-Weiss temperature regime ($\gtrsim 100$~K for \ecp\, where the AFM spin correlations are sufficiently weak), the $\chi_J(T)$ is isotropic $[\chi_{Jab}(T)=\chi_{Jc}(T)]$, so Eq.~(\ref{Eq:Delta1chi}) becomes
\be
\frac{1}{\chi_c(T)} - \frac{1}{\chi_{ab}(T)} = \frac{3}{2}A_c,
\label{Eq:Delta1chi2}
\ee
where the difference on the left-hand side is thus temperature independent.  Shown in Fig.~\ref{Fig:Chi_1T_EuCo2P2_TN66_ka88}(a) is a plot of $\chi_c^{-1}(T) - \chi_{ab}^{-1}(T)$ versus~$T$ obtained from the data for $H=1$~T in Fig.~\ref{Fig:EuCo2P2_InvChi}(a).  One indeed sees that the difference is nearly independent of~$T$ above 100~K with an average value $\frac{3}{2}A_c = 0.584(6)\,{\rm mol/cm^3}$, yielding
\be
A_c = 0.389(4)\,{\rm \frac{mol}{cm^3}}, \qquad A_{ab} = -0.195(2)\,{\rm \frac{mol}{cm^3}},
\label{Eq:LambdaVals}
\ee
where we used Eq.~(\ref{Eq:lambdaSum}) to obtain the latter equality.

Using the experimental $\chi_{\alpha}$ data in Fig.~\ref{Fig:EuCo2P2_chi}(a) and Eqs.~(\ref{Eq:ChiJalpha}) and~(\ref{Eq:LambdaVals}), we obtained the Heisenberg contributions $\chi_{Jc}(T)$ and $\chi_{Jab}(T)$ shown in Fig.~\ref{Fig:Chi_1T_EuCo2P2_TN66_ka88}(b).  For $T\lesssim 100$~K, $\chi_{Jab}$ starts to become less than $\chi_{Jc}$ due to dynamic short-range AFM ordering above $T_{\rm N}$\@.  Below $T_{\rm N} = 66.5$~K, $\chi_{Jc}$ becomes nearly independent of~$T$ as expected from MFT for a $c$-axis helix \cite{Johnston2012, Johnston2015}.

The normalized $\chi_{Jab}(T \leq T_{\rm N})/\chi_J(T_{\rm N})$ for a helical Heisenberg AFM is given by \cite{Johnston2012,Johnston2015}
\bse
\label{Eqs:Chixy}
\begin{equation}
\frac{\chi_{Jab}(T \leq T_{\rm N})}{\chi_J(T_{\rm N})}=  \frac{(1+\tau^*+2f+4B^*)(1-f)/2}{(\tau^*+B^*)(1+B^*)-(f+B^*)^2},
\label{eq:Chi_plane}
\end{equation}
where
\begin{equation}
B^*=  2(1-f)\cos(kd)\,[1+\cos(kd)] - f,
\label{eq:Bstar}
\end{equation}
\be
t =\frac{T}{T_{\rm N}},\quad \tau^*(t) = \frac{(S+1)t}{3B'_S(y_0)}, \quad y_0 = \frac{3\bar{\mu}_0}{(S+1)t}, 
\ee
\ese
the ordered moment versus $T$ in $H=0$ is denoted by $\mu_0$, the reduced ordered moment $\bar{\mu}_0 = \mu_0/\mu_{\rm sat}$ is determined by solving
$\bar{\mu}_0 = B_S(y_0)$, $B'_S(y_0) = [dB_S(y)/dy]|_{y=y_0}$ and our unconventional definition of the Brillouin function $B_S(y)$ is given in Refs.~\onlinecite{Johnston2012} and~\onlinecite{Johnston2015}.

We fitted the $\chi_{Jab}(T)$ data in Fig.~\ref{Fig:Chi_1T_EuCo2P2_TN66_ka88}(b) by Eqs.~(\ref{Eqs:Chixy}) using \mbox{$S=7/2$} and $f = 0.346$, which is slightly larger than the value $f_{\rm ave} = 0.320$ in Table~\ref{Table: curieFit}.   For $kd(T)$ we used the neutron diffraction value $kd(T=64~{\rm K}) = 0.834\pi$ from Eq.~(\ref{Eq:kdNeuts}) \cite{Reehuis1992}.  In order to fit the lowest-$T$ data in Fig.~\ref{Fig:Chi_1T_EuCo2P2_TN66_ka88}, we used $kd(T=0) = 0.88\pi$, which is comparable to the value at 15~K in Eq.~(\ref{Eq:kdNeuts}) \cite{Reehuis1992}.  For intermediate temperatures we linearly interpolated $kd$ between these two values.  The $\chi_{ab}(T\leq T_{\rm N})$ thus obtained from our MFT is plotted as the solid blue curve in Fig.~\ref{Fig:Chi_1T_EuCo2P2_TN66_ka88}(b).  The $T$~dependence of the fit is seen to be in excellent agreement with the data.

\section{\label{Sec:HeisExchInts} Heisenberg Exchange Interactions}

We now estimate the intralayer and interlayer Heisenberg exchange interactions within the minimal $J_0$-$J_{1z}$-$J_{2z}$ MFT model for a helix in Fig.~\ref{Fig:J0_Jz1_Jz2_model_helix} \cite{Nagamiya1967}, where $J_0$ is the sum of all Heisenberg exchange interactions of a representative spin to all other spins in the same spin layer perpendicular to the helix ($c$) axis, $J_{1z}$ is the sum of all interactions of the spin with spins in an adjacent layer along the helix axis, and $J_{2z}$ is the sum of all interactions of the spin with spins in a second-nearest layer.  Within this model $kd$, $T_{\rm N}$ and $\theta_{{\rm p}}$ are related to these exchange interactions by \cite{Johnston2012, Johnston2015}
\bse
\label{Eqs:J0J1zJ2z}
\bea
&&\cos(kd) = -\frac{J_{z1}}{4J_{z2}},\\*
T_{\rm N} &=& -\frac{S(S+1)}{3k_{\rm B}} \big[J_0 + 2J_{z1}\cos(kd)\nonumber\\*
&& \hspace{0.9in} +\ 2J_{z2}\cos(2kd)\big], \label{eq:TN}\\*
\theta_{\rm p} &=& -\frac{S(S+1)}{3k_{\rm B}} \left(J_0+2J_{z1}+2J_{z2}\right),
\label{eq:thetap}
\eea
\ese
where a positive (negative) $J$ corresponds to an AFM (FM) interaction.  Using $S = 7/2,\ T_{\rm N} = 66.5\ {\rm K},\ \theta_{\rm p}=\theta_{\rm p\,ave}=21.4$~K, the average $kd=0.857\pi$ of the two $kd$ values in Fig.~\ref{Fig:Chi_1T_EuCo2P2_TN66_ka88}(b) and solving Eqs.~(\ref{Eqs:J0J1zJ2z}) for the three exchange constants, one obtains
\bea
J_0/k_{\rm B} &=& -9.55~{\rm K},\quad J_{z1}/k_{\rm B} = 2.14~{\rm K}, \label{Eq:J0Jz1Jz2Vals}\\*
&&J_{z2}/k_{\rm B} = 0.594~{\rm K},\nonumber
\eea
where the variation in $kd$ with $T$ for $0<T<T_{\rm N}$ is found to have a minimal effect on the derived $J$'s.  As anticipated in Sec.~\ref{Sec:Intro}, from the FM-like $\theta_{\rm p}$ the net exchange constant $J_0+2J_{z1}+2J_{z2} = -4.08$~K is FM, and the out-of-plane exchange constants $J_{z1}$ and $J_{z2}$ are both AFM.

\begin{figure}
\includegraphics[width=1.25in]{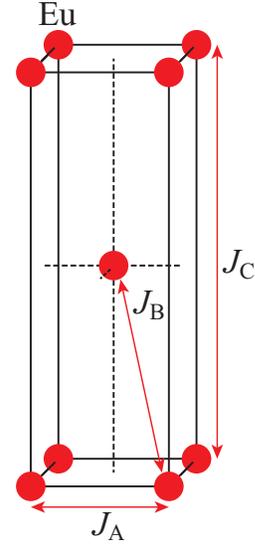}
\caption {(Color online) Body-centered Eu sublattice, where $c/a = 3$.  The Heisenberg exchange interactions $J_{\rm A},\ J_{\rm B}$ and $J_{\rm C}$ are defined in the figure.}
\label{Fig:bct_Eu_lattice}
\end{figure}

\begin{table}
\caption{\label{Table:EuCo2P2_Js} Exchange constants between Eu spins $J_{\rm A}$, $J_{\rm B}$ and $J_{\rm C}$ in Fig.~\ref{Fig:bct_Eu_lattice} determined by fitting the $\chi_{ab}(T\leq T_{\rm N})$ data by MFT and from electronic band structure (EBS) calculations for Eu spins $S = 7/2$ using Eqs.~(\ref{Eqs:JAJBJCESC} ). Negative $J$ values are FM and positive values are AFM\@.  Also shown are the Weiss temperatures $\theta_{\rm p}$ in the Curie-Weiss law~(\ref{Eq:CW-law}) calculated from Eq.~(\ref{Eq:thetapCalc}) and the listed $J$ values.  The MFT value of $\theta_{\rm p}$ is equal to the observed value by construction.}
\begin{ruledtabular}
\begin{tabular}{lccccc}
method 	& $J_{\rm A}/k_{\rm B}$	& $J_{\rm B}/k_{\rm B}$  	& $J_{\rm C}/k_{\rm B}$	& 	$J_{\rm B}/J_{\rm C}$ & $\theta_{\rm p}$\\
	& (K) & (K) & (K) & (K) & (K)\\
\hline 
MFT  		&  $-2.39$ 	&  0.535 	&  0.594	& 0.90	& 21.5\\
EBS  			&   	&   	&   		\\
$S_{\rm eff}^2 = S(S+1)$ 	& $-2.20$  & 1.35  	& 1.49  	& 0.91 & $-26.1$	\\
$S_{\rm eff}^2 = S^2$ 		& $-2.82$  & 1.74  	& 1.92  	& 0.91 & $-20.3$	\\
\end{tabular}
\end{ruledtabular}
\end{table}

The bct Eu sublattice of \ecp\ is shown in Fig.~\ref{Fig:bct_Eu_lattice}, where the measured ratio $c/a \approx 3$ is to scale.  Assuming that the exchange interactions $J_{\rm A}$, $J_{\rm B}$ and $J_{\rm C}$ in the figure are the only ones present, in terms of the interactions in the $J_0$-$J_{z1}$-$J_{z2}$ model one has
\be
J_0 = 4J_{\rm A}, \quad J_{z1} = 4J_{\rm B}, \quad J_{z2} = J_{\rm C}.
\ee
Then using Eq.~(\ref{Eq:J0Jz1Jz2Vals}) one obtains the MFT $J$ values listed in Table~\ref{Table:EuCo2P2_Js}.

The Hamiltonian associated with a single spin in a spin system in $H=0$ with no anisotropy and containing identical crystallographically-equivalent spins is
\be
{\cal H}_i = \frac{1}{2} \sum_j J_{ij}{\bf S}({\bf R}_i)\cdot{\bf S}({\bf R}_j),
\label{Eq:Hi}
\ee
where the factor of 1/2 arises because the energy of an interacting spin pair is equally shared between the two spins in the pair, the sum is over the neighbors ${\bf S}({\bf R}_j)$ of the given central spin ${\bf S}({\bf R}_i)$ and the $J_{ij}$ are the Heisenberg exchange interactions between each respective spin pair.  
Here we only consider Bravais spin lattices where the position of each spin is a position of inversion symmetry of the spin lattice such as the body-centered-tetragonal (bct) spin lattice in Fig.~\ref{Fig:bct_Eu_lattice}.  We further restrict our discussion to coplaner AFMs in which the ordered moments in the ordered AFM state are aligned in the $xy$ plane.

The expression for the classical ground-state energy per spin obtained from Eq.~(\ref{Eq:Hi}) is
\be
E_i = \frac{S^2}{2} \sum_j J_{ij}\cos\phi_{ji},
\label{Eq:Ei}
\ee
where $\cos\phi_{ji} = \hat{\bf S}({\bf R}_i)\cdot\hat{\bf S}({\bf R}_j)$ and $\phi_{ji}$ is the azimuthal angle within the $xy$~plane between ${\bf S}({\bf R}_j)$ and ${\bf S}({\bf R}_i)$.  One can write $\phi_{ji}$ as
\be
\phi_{ji} = {\bf Q}\cdot {\bf R}_{ji},
\ee
where ${\bf R}_{ji} = {\bf R}_j - {\bf R}_i$.  Then for a system containing $N$~spins Eq.~(\ref{Eq:Ei}) becomes
\bse
\bea
E({\bf Q}) = \frac{NS^2}{2} \sum_j J_{ij}\cos\left({\bf Q}\cdot{\bf R}_{ji}\right) = \frac{NS^2}{2}J({\bf Q}),
\label{Eq:E0(Q)}
\eea
where
\be
J({\bf Q}) = \sum_j J_{ij}\cos\left({\bf Q}\cdot{\bf R}_{ji}\right)
\label{Eq:JofQ}
\ee
\ese
is the cosine Fourier transform of the position-dependent exchange interaction.  Thus the {\bf Q} with the lowest algebraic $J({\bf Q})$ is the classical ground-state AFM ordering wave vector.

In the present system \ecp\ with a helical ground state, one has
\bse
\label{Eqs:QComps}
\be
{\bf Q} = (0,0,k_{\rm ave}),
\ee
where from Fig.~12(b), the average $k$ between $T=0$ and $T_{\rm N}$ is
\be
Q_z = k_{\rm ave} = 0.857\frac{2\pi}{c}.
\label{Eq:Qzk}
\ee
\ese
From Fig.~\ref{Fig:bct_Eu_lattice}, Eq.~(\ref{Eq:JofQ}) yields
\bea
J({\bf Q}) &=& J_{\rm A} \sum_{j=1}^4\cos({\bf Q}\cdot {\bf R}_{{\rm A}ji}) + J_{\rm B} \sum_{j=1}^8\cos({\bf Q}\cdot {\bf R}_{{\rm B}ji})\nonumber\\
&& \hspace{0.7in} +\ J_{\rm C} \sum_{j=1}^2\cos({\bf Q}\cdot {\bf R}_{{\rm C}ji})
\eea
where the first, second and third sums are over spin neighbors ${\bf S}_j$ connected to ${\bf S}_i$ by interactions $J_{\rm A}$, $J_{\rm B}$ and $J_{\rm C}$, respectively.  From the spin positions in Fig.~\ref{Fig:bct_Eu_lattice} one obtains
\bea
J({\bf Q})&=& 2\Bigg\{J_{\rm A}[\cos(Q_x a)+\cos(Q_y a)]\label{Eq:JofQECP}\\
&& +\ 4J_{\rm B}\cos\left(\frac{Q_x a}{2}\right)\cos\left(\frac{Q_y a}{2}\right)\cos\left(\frac{Q_z c}{2}\right)\nonumber\\
&& +\ J_{\rm C}\cos\left(Q_z c\right)\Bigg\}\nonumber
\eea
Inserting ${\bf Q} = (0,0,Q_z)$ with $Q_z$ from Eq.~(\ref{Eq:Qzk}) into Eq.~(\ref{Eq:JofQECP}) and using the exchange interactions in Table~\ref{Table:EuCo2P2_Js} gives $J({\bf Q})/k_{\rm B} = -12.7$~K\@.  Hence Eq.~(\ref{Eq:E0(Q)}) gives the ground state energy of the helix in \ecp\ with $S=7/2$ as
\be
\frac{E[{\bf Q} = (0,0,k_{\rm ave})]}{Nk_{\rm B}} = -{\rm 78~K}.
\ee
The magnitude of this quantity is of order $T_{\rm N}=66.5$~K for this compound.

\section{\label{Sec:SpinWaves} Spin Waves and Magnetic Heat Capacity at Low Temperatures}

\begin{figure}
\includegraphics[width=3.3in]{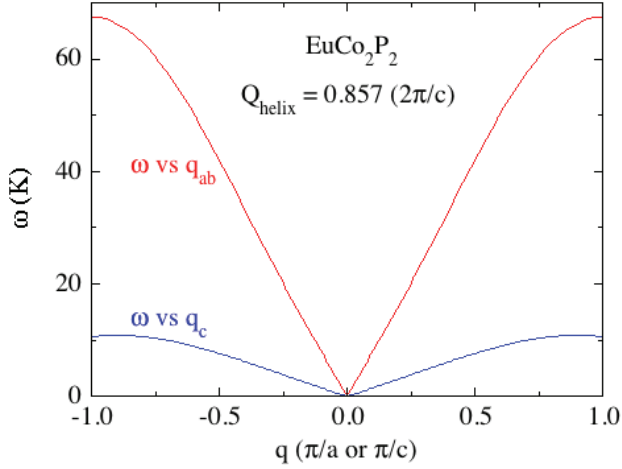}
\caption {(Color online) Spin-wave dispersion relation $\omega$ versus $q$ in the $ab$ plane and along the $c$~axis for the helix wave vector given in the figure.}
\label{Fig:OmegaOFqS72}
\end{figure}

The value of $\beta$ in Eq.~(\ref{Eq:Cp_Fit}) describing the $T^3$ contribution to the low-$T$ heat capacity of \ecp\ is found to be too large to arise from lattice vibrations.  We therefore infer that the excess contribution is due to thermal excitations of three-dimensional AFM spin waves in the helix with a $T^3$ contribution.  To calculate that contribution, we first calculate the spin wave dispersion relation $\omega({\bf q})$ with wavevector ${\bf q} = (q_x,q_y,q_z)$ propagating along the $x=a$, $y=b$ and $z=c$ axes.  In the absence of an anisotropy energy gap, the nondegenerate spin-wave dispersion relation for a helical AFM ground state is given from linear spin-wave theory as  \cite{Nagamiya1967}
\bea
\hbar \omega({\bf q}) &=& S\Bigg\{2\left[J({\bf Q})-\frac{1}{2}\left[J({\bf Q}+{\bf q}) + J({\bf Q}-{\bf q})\right]\right]\nonumber\\
&& \hspace{0.75in}\times\left[J({\bf Q}) - J({\bf q})\right]\bigg\}^{1/2},
\label{Eq:OmegaOFq}
\eea
where $\hbar$ is Planck's constant divided by $2\pi$ and the vector function $J({\bf x})$ is given in Eq.~(\ref{Eq:JofQECP}).  Plots of $\omega({\bf q})$ about {\bf q} = 0 for spin waves propagating in the $ab$~plane and along the $c$~axis are shown in Fig.~\ref{Fig:OmegaOFqS72} obtained using the helix wave vector in Eqs.~(\ref{Eqs:QComps}), Eqs.~(\ref{Eq:JofQECP}) and~(\ref{Eq:OmegaOFq}), the MFT $J$ values in Table~\ref{Table:EuCo2P2_Js}, and the lattice parameters in Table~\ref{Table:EuCo2P2_xrd parameter}.

One observes from Fig.~\ref{Fig:OmegaOFqS72} that $\omega \propto q$ for small $q$ in each of the principal axis directions as expected for AFM spin waves in the absence of an anisotropy gap.  In this limit of small~$q$ one has
\be
\omega = \sqrt{v_{ab}^2(q_a^2 + q_b^2) + v_c^2 q_c^2},
\ee
where $v_{ab}$ and $v_c$ are the spin-wave velocities (speeds) in the $ab$~plane and along the $c$~axis, respectively.  For a helix wave vector ${\bf Q} = (0,0,Q_z)$ where $Q_z>0$, Eqs.~(\ref{Eq:JofQECP}) and~(\ref{Eq:OmegaOFq}) yield
\bse
\label{Eqs:vabc}
\bea
v_{ab} &=& \frac{4aSk_{\rm B}}{\hbar}\sin\bigg(\frac{Q_z c}{4}\bigg)\Bigg\{-\bigg[J_{\rm A}+J_{\rm B}\cos\left(\frac{Q_zc}{2}\right)\bigg]\nonumber\\
&& \times\ \bigg[2J_{\rm B} + J_{\rm C} + J_{\rm C}\cos\bigg(\frac{Q_zc}{2}\bigg)\bigg]\Bigg\}^{1/2}
\eea
and
\bea
v_c &=& \frac{4cSk_{\rm B}}{\hbar}\sin\left(\frac{Q_z c}{4}\right)\\
&&\times\Bigg\{-\bigg[J_{\rm B}\cos\left(\frac{Q_z c}{2}\right)  + J_{\rm C} \cos(Q_z c)\bigg]\nonumber\\
&& \hspace{0.3in}\times\ \bigg[2J_{\rm B}+J_{\rm C}+J_{\rm C}\cos\left(\frac{Q_zc}{2}\right)\bigg]\Bigg\}^{1/2}.\nonumber
\eea
\ese
Inserting $Q_z$ from Eq.~(\ref{Eq:Qzk}), the MFT $J_\alpha$ values from Table~\ref{Table:EuCo2P2_Js} and the lattice parameters from Table~\ref{Table:EuCo2P2_xrd parameter} into Eqs.~(\ref{Eqs:vabc}) yields
\be
v_{ab} = 1.21\times10^3~{\rm m/s}, \quad v_c = 0.72\times10^3~{\rm m/s}.
\label{Eq:vabc}
\ee
The spin-wave velocity is thus 70\% larger in the $ab$~plane than along the $c$~axis.

The contribution of the spin waves to the low-$T$ molar heat capacity of \ecp\ is $C_{\rm SW} = \beta_{\rm SW} T^3$, where $\beta_{\rm SW}$ is obtained from Eq.~(73) in Ref.~\onlinecite{Johnston2011} as 
\be
\beta_{\rm SW} = \frac{\pi^2Rk_{\rm B}^3V_{\rm spin}}{15\hbar^3 v_{ab}^2v_c},
\ee
which is a factor of four smaller than in Eq.~(73) in Ref.~\onlinecite{Johnston2011} due to a fourfold reduction in the spin-wave degeneracy in the present case.  Here we use SI units and $V_{\rm spin} = a^2c/2$ is the volume per spin, yielding
\be
\beta_{\rm SW} = 0.92~{\rm \frac{mJ}{mol\,K^4}}.
\ee
This is a significant fraction of $\beta = 2.8~{\rm mJ/(mol\,K^4)}$ measured for \ecp\ as given in Table~\ref{Table:Heat_capacity}.  Thus the Debye temperature of 151~K obtained from the $\beta$ value for \ecp\ as listed in Table~\ref{Table:Heat_capacity} is substantially underestimated.  Furthermore, in view of the small energy width of $\omega(q_c)$ in Fig.~\ref{Fig:OmegaOFqS72}, the spin-wave contribution to $C_{\rm p}$ at low~$T$ is likely larger than calculated here.

\section{\label{Sec:ElecStruct} Electronic Structure Calculations}

In order to check the accuracy of the MFT model for EuCo$_2$P$_2$ and for a more complete understanding of this interesting system, we performed {\it ab~initio} total energy and electronic band structure calculations for several spin structures.  Density functional theory (DFT) is so far the only approximation which allows such simulations for generic macroscopic systems within a reasonable computing time. DFT has some drawbacks since the nonlocal part of the exchange interaction is not explicitly considered \cite{DMFT}. However, it has been found to give rather good estimates of electronic properties, far from metal-insulator transitions \cite{MITrev}. In this work, we employed the DFT implementation known as Dmol$^3$ \cite{Delley96,Weinert92,Delley98}. Dmol$^3$ offers the combination of  Becke's exchange \cite{Becke88} with the One-parameter Progressive correlation-functional \cite{Tsuneda99}, which provides a better representation of equal- and opposite-spin correlations (in comparison with the standard Perdew-Burke-Ernzerhof functional). In addition, Dmol$^3$ uses a linear combination of atomic orbitals as basis set and therefore does not require the replacement of the true lattice potential with pseudopotentials. It also allows us to treat core and valence electrons on an equal basis, including scalar-relativistic corrections. 

\subsection{Heisenberg Exchange Interactions}

\begin{figure}
\includegraphics[width=3in]{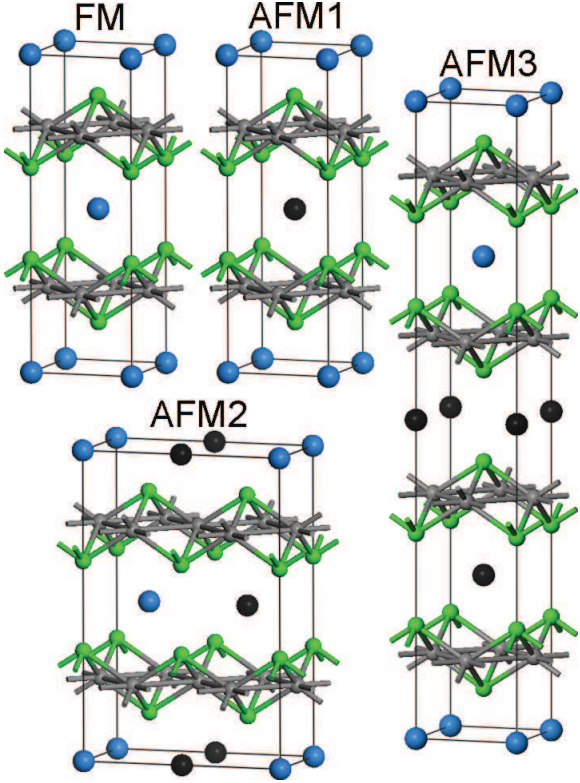}
\caption {(Color online) Magnetic unit cells of ferromagnetic (FM) and three antiferromagnetic (AFM) ordered Eu spin lattices for which the total energies were calculated.  The solid blue and black spheres correspond within these collinear structures to Eu spins up and spins down, respectively.  The solid green circles represent As atoms and the smaller solid  brown circles represent Co atoms.  The collinear ordering axis of the Eu spins is arbitrary with respect to the present DFT calculations.}
\label{Fig:spinsymmetries}
\end{figure}

Binding energies of EuCo$_2$P$_2$ with four symmetry-inequivalent spin structures are needed for the  calculation of the three isotropic exchange constants $J_{\rm A}$, $J_{\rm B}$, and $J_{\rm C}$ in Fig.~\ref{Fig:bct_Eu_lattice}.  We took the FM configuration (spin lattice symmetry group $I4/mmm$, No.~139, tetragonal), and three AFM structures (in the following referred as AFM1, AFM2 and AFM3)  corresponding to ordered spin-lattice symmetries in the respective space groups $P4/mmm$, No.~123, tetragonal; $Pmmn$, No.~59, orthorhombic; and $P4/nmm$, No.~129, tetragonal.  In AFM1, the spin at the center of the body-centered tetragonal magnetic unit cell is oriented opposite to the spins on the cell corners.  AFM2 is obtained from the chemical cell by doubling either the $a$ or the $b$ axis and inverting the image spins. AFM3 is obtained similarly to AFM2, but doubling the cell along the $c$ axis. A picture of these ordered-moment structures is shown in Fig.~\ref{Fig:spinsymmetries}.  Since the calculations only include scalar-relativistic corrections, spin-orbital effects are not included.  Hence the easy ordering axis of the Eu spins is undetermined for the FM and AFM structures in Fig.~\ref{Fig:spinsymmetries}.  The mapping of the total energies per Eu onto a set of exchange constants is achieved with the set of equations
\begin{subequations}
\label{Eqs:EFMAFM}
\begin{eqnarray}
E_{\rm FM} &=& E_0+S^2_{\text{eff}}\left(2J_{\rm A}+4J_{\rm B}+J_{\rm C}\right) \label{EnFM}\\
E_{\rm AFM1} &=& E_0+S^2_{\text{eff}}\left(2J_{\rm A}-4J_{\rm B}+J_{\rm C}\right)  \label{EnAF1}\\
E_{\rm AFM2} &=& E_0+S^2_{\text{eff}} J_{\rm C} \label{EnAF2} \\
E_{\rm AFM3} &= &E_0+S^2_{\text{eff}}\left(2J_{\rm A}-J_{\rm C}\right), \label{EnAF3}
\end{eqnarray}      
\end{subequations}
where $S^2_{\text{eff}}$ is related (but not necessarily equal) to the square of the spin quantum number.

To ensure the accuracy of our results we performed several runs for  each spin configuration with increasing resolution in ${\bm k}$~space, starting with a distance of $|\Delta {\bm k}|<0.04\,\text{\AA}^{-1}_{}$ between ${\bm k}$ points. We found that already for $|\Delta {\bm k}|=0.025\,\text{\AA}^{-1}$ and $|\Delta {\bm k}|=0.015\,\text{\AA}^{-1}$ the differences in total energy could be discounted; i.e., these were all less than $10^{-4}$\,eV\@. The magnetic interactions are two orders of magnitude larger and therefore the calculated exchange constant are numerically precise within 1\%.

Since the squared spin is  not a conserved quantity in general within DFT calculations of periodic quantum systems,  the determination of exchange constants from total energies is (in principle) an ill-posed problem. It is not clear which value should be  assigned to the effective constant $S^2_{\text{eff}}$ in Eqs.~(\ref{Eqs:EFMAFM}).  Therefore, we made estimates taking $S^2_{\text{eff}}$ equal to the classical $S^2_{}$ and the quantum $S(S+1)$ values; however, it should be noted that these are just bounds for the orders of magnitude and signs. The calculated local spin of an Eu$^{+2}$ ion was, in all structures,  equivalent to the theoretical value ($S=7/2$) within 1\%, in agreement with the present and previous \cite{Morsen1988, Reehuis1992} experimental results. The polarization of Co and P ions was also found to be negligible in all four spin structures, which supports the use of the Heisenberg model (with localized Eu$^{+2}$ spins) and agrees with previous conclusions that the Co atoms do not contribute to the observed helical AFM structure of \ecp\ \cite{Morsen1988, Reehuis1992}. 

The calculated total energies satisfy the relation
\begin{equation}
E_{\rm AFM1} < E_{\rm AFM3}<E_{\rm AFM2}<E_{\rm FM}\,  .
\end{equation}       
\noindent AFM1 is, among the chosen collinear configurations, the closest  to the observed noncollinear helical ground state and the energy differences per Eu atom for the other structures are
\bse
\bea
(E_{\rm AFM3}-E_{\rm AFM1})/k^{}_{\text{B}} &=& 38.1\,\text{K},\\*
(E_{\rm AFM2}-E_{\rm AFM1})/k^{}_{\text{B}} &=& 154\,\text{K},\\*
(E_{\rm FM}-E_{\rm AFM1})/k^{}_{\text{B}} &=& 170\,\text{K}.
\eea
\ese
The resulting  exchange constants are
\bse
\label{Eqs:JAJBJCESC}
\bea
S^2_{\text{eff}}J_{\rm A}/k^{}_{\text{B}} &=& -34.6\,\text{K},\\*
S^2_{\text{eff}}J_{\rm B}/k^{}_{\text{B}} &=& 21.3\,\text{K},\\*
S^2_{\text{eff}}J_{\rm C}/k^{}_{\text{B}} &=& 23.5\,\text{K}.
\eea
\ese
Thus, interactions in the $ab$ plane are predicted to be FM (negative sign of $J_{\rm A}$) and the interplane ones are AFM, in  agreement with the values in Table~\ref{Table:EuCo2P2_Js} obtained from the MFT fit of the experimental $\chi_{ab}(T\leq T_{\rm N})$ data. Furthermore, the calculated ratio $J_{\rm B}/J_{\rm C}=0.91$ is very close to the value of $0.90$ obtained from the MFT fit.  This ratio determines the turn angle of the helix between adjacent FM-aligned planes of spins perpendicular to the helix wave vector and is therefore also in excellent agreement with the observed \cite{Reehuis1992} AFM ground state.

Concerning the absolute $J$ values, we find that the $J_{\rm A}$ values computed from Eqs.~(\ref{Eqs:JAJBJCESC}) using $S_{\rm eff}^2 = S(S+1)$ and $S^2$ are similar to the MFT-fitted result as shown in Table~\ref{Table:EuCo2P2_Js}.  However, the calculated interplane constants   ($J_{\rm B}$ and $J_{\rm C}$) are about twice as large as the MFT-fitted ones as also shown in Table~\ref{Table:EuCo2P2_Js}.  For instance, $J_{\rm B}/k^{}_{\text{B}}=1.35\,\text{K}$ for $S_{\rm eff}^2 = S(S+1)$ while the MFT-fitted value $0.535\,\text{K}$ is $2.5$ times smaller. These differences may have various origins.  One possibility is the inexact representation of exchange interactions within DFT\@.  Another is the neglect of spin fluctuations within mean-field models, which results in the need of smaller exchange constants for a given critical temperature \cite{DMFT,MITrev}. Thus, it is likely that the actual magnetic interactions are stronger that those obtained from the MFT fit.  Yet another possibility is that Heisenberg interactions other than those shown in Fig.~\ref{Fig:bct_Eu_lattice} are important.  All interactions between spins within a layer and between spins in a layer and those in first- and second-neighbor layers are all taken into account in the $J_0$-$J_{z1}$-$J_{z2}$ MFT model, but not when assigning these interactions to only $J_{\rm A}$, $J_{\rm B}$ and $J_{\rm C}$ for both the MFT model and the DFT calculations.

A diagnostic for the last possibility is a calculation of the Weiss temperature $\theta_{\rm p}$ in the Curie-Weiss law~(\ref{Eq:CW-law}) using the band-structure exchange constants and comparing the result with experiment.  Referring to Fig.~\ref{Fig:bct_Eu_lattice}, within the $J_{\rm A}$-$J_{\rm B}$-$J_{\rm C}$ Heisenberg model one has \cite{Johnston2012, Johnston2015}
\bea
\theta_{\rm p} &=& -\frac{S(S+1)}{3}\sum_j \frac{J_{ij}}{k_{\rm B}} \nonumber\\*
&=& -\frac{S(S+1)}{3}\left(4\frac{J_{\rm A}}{k_{\rm B}}+8\frac{J_{\rm B}}{k_{\rm B}}+2\frac{J_{\rm C}}{k_{\rm B}}\right),
\label{Eq:thetapCalc}
\eea
where the sum is over neighors~$j$ of a representative central spin~$i$.  The results of this calculation for both the MFT and DFT values of $J_{\rm A}$, $J_{\rm B}$ and $J_{\rm C}$ are shown in Table~\ref{Table:EuCo2P2_Js}.  The MFT value of +21.4~K agrees with experiment by construction.  However, one sees that the $J$ values from the DFT calculations lead to $\theta_{\rm p} = -26$~K and $-20$~K for $S_{\rm eff}=S(S+1)$ and $S^2$, respectively, which are both of the correct magnitude but of the wrong (negative) sign, suggesting dominant AFM interactions rather than dominant FM interactions as required from the observed value of $\theta_{\rm p}$.  This disagreement in turn suggests that additional exchange interactions between the Eu spins beyond $J_{\rm A}$, $J_{\rm B}$ and $J_{\rm C}$ are present in \ecp.  For instance, FM exchange interactions among next-nearest neighbors within the $ab$ plane (diagonals) presumably play a role.  The addition of this new ($J_{\rm D}$) coupling to Eqs.~(\ref{Eqs:EFMAFM}) does not alter the ratio $J_{\rm B}/J_{\rm C}$ and increases the absolute value of the $J_{\rm A} + J_{\rm D}$ sum. The numerical determination of $J_{\rm D}$ would require the calculation of the total energy for another spin structure but it can already be inferred from Eqs.~(\ref{Eqs:EFMAFM}) that it would bring the calculated $theta_{\rm p}$ closer to the experimental value.

\subsection{Electronic Band Structure}

\begin{figure}
\includegraphics[width=0.5\linewidth]{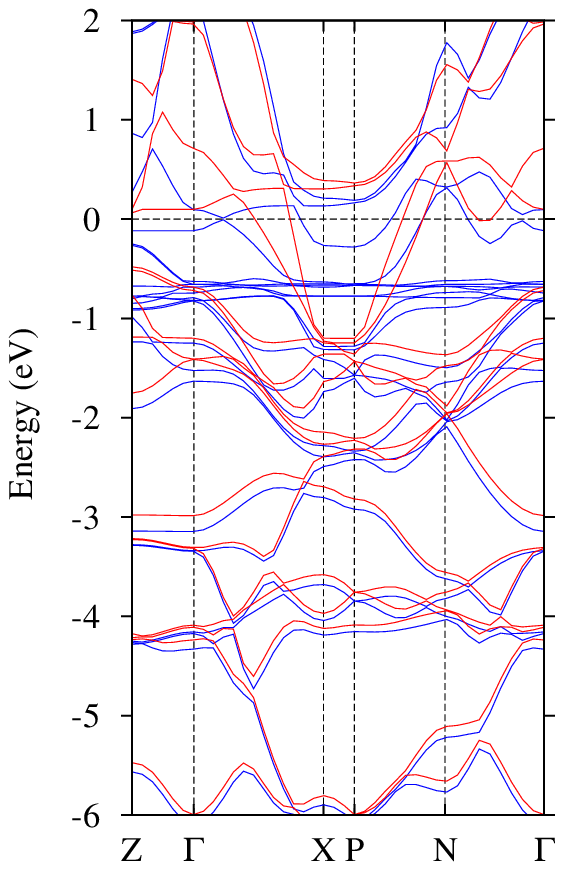}\includegraphics[width=0.5\linewidth]{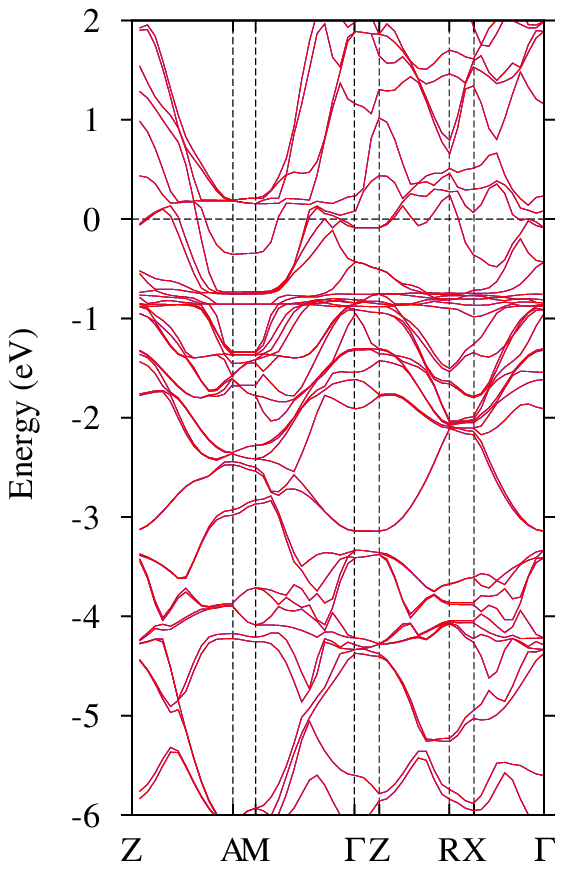}\\
\includegraphics[width=0.5\linewidth]{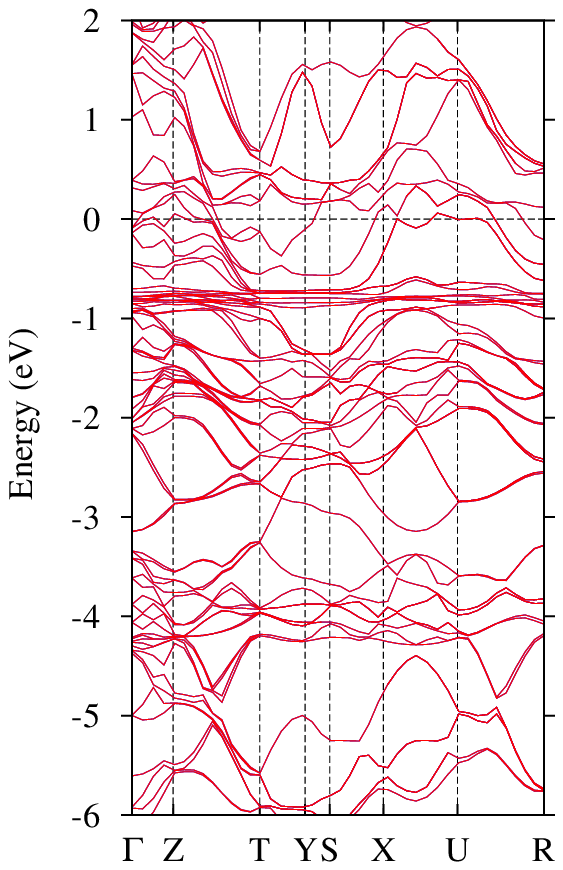}\includegraphics[width=0.5\linewidth]{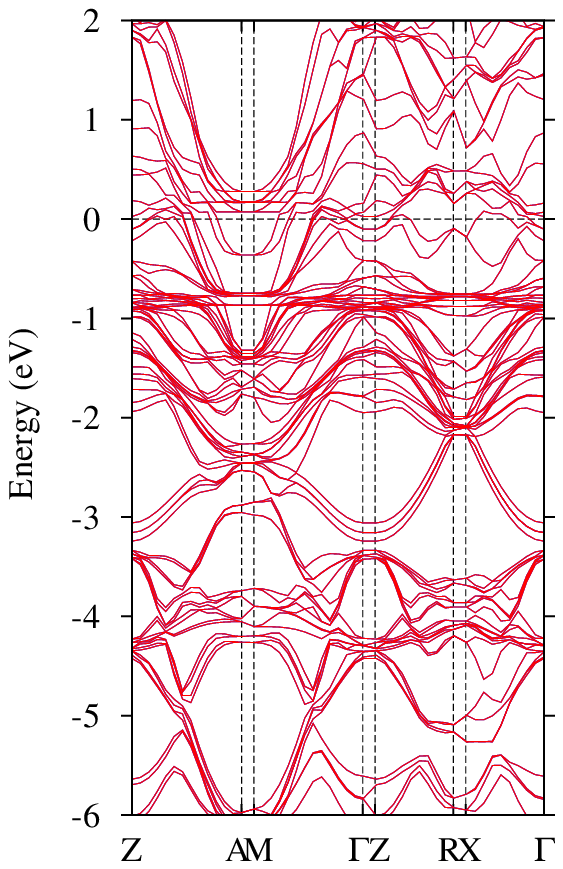}
\caption{Electronic band structure for the calculated ordered spin structures. Upper left, FM; upper right AFM1; lower left, AFM2; and lower right, AFM3. Spin up (down) states are drawn in blue (red).  For the AFM figures, the blue curves are largely obscured by the red ones.}
\label{fig:1}
\end{figure}

\begin{figure}
\includegraphics[width=0.5\linewidth]{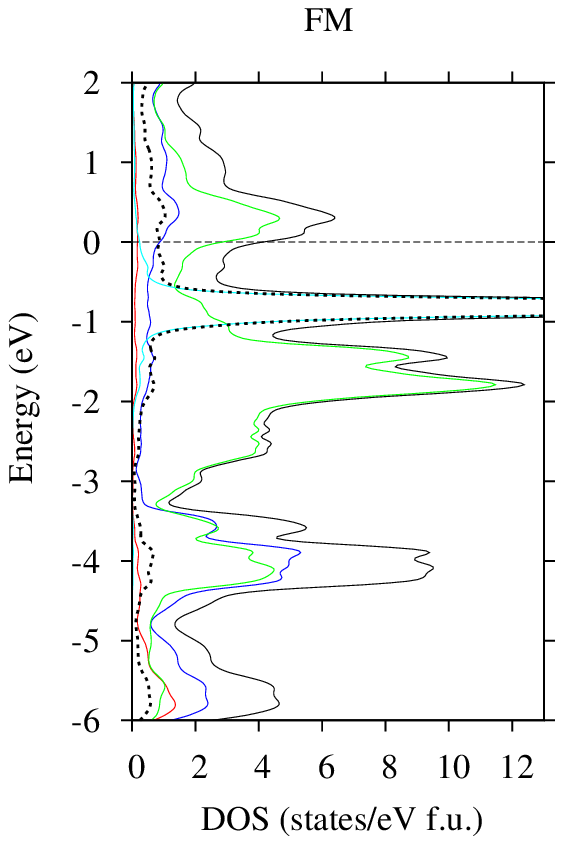}\includegraphics[width=0.5\linewidth]{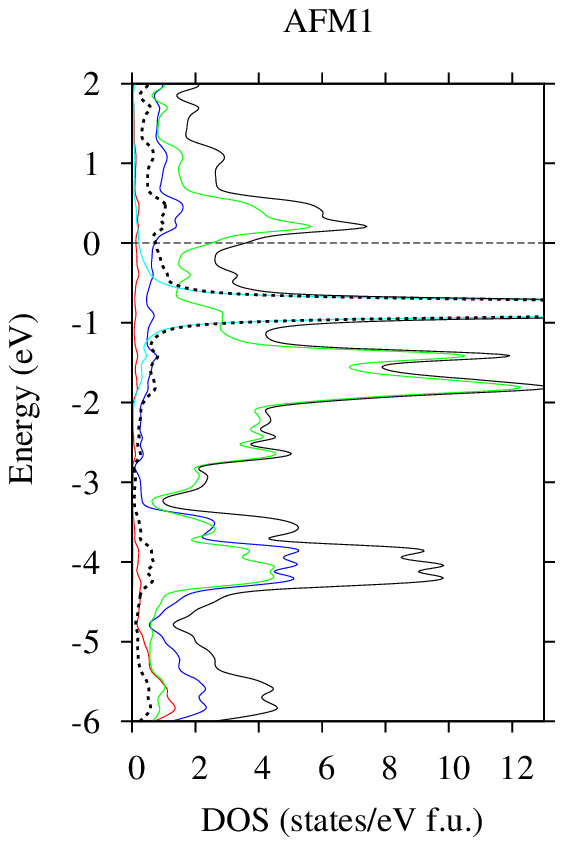}\\
\includegraphics[width=0.5\linewidth]{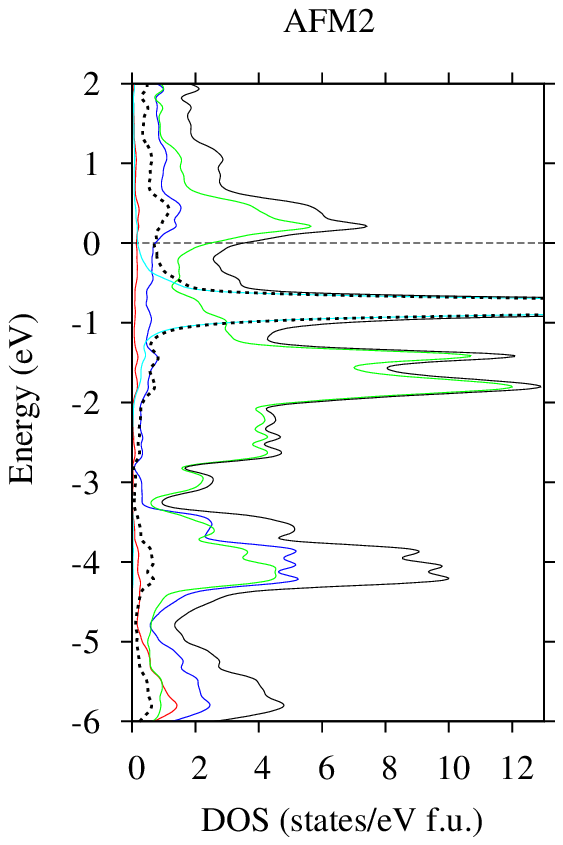}\includegraphics[width=0.5\linewidth]{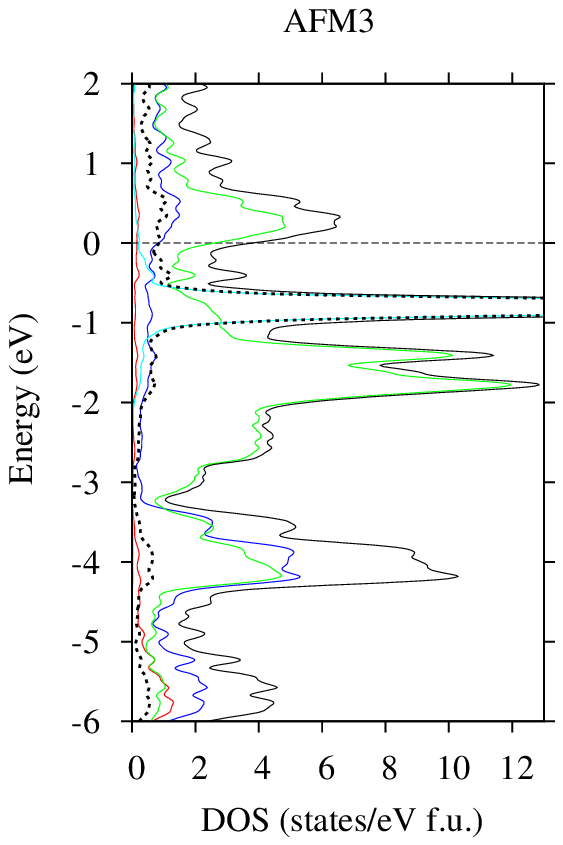}
\caption{Electronic partial density of states (PDOS), showing $s,\ p,\ d$ and $f$ contributions in red, blue, green and cyan, respectively, for the four spin structures for which LDA calculations were done.  Also shown is the total DOS (black) and the total contributions of the Eu atoms (black thick dotted curves).  The zero of energy is $E_{\rm F}$ (horizontal dotted lines). }
\label{fig:2}
\end{figure}

The electronic band structures for the calculated spin symmetries are shown in Fig.~\ref{fig:1} and the corresponding contributions of~$d$ and~$p$ orbitals to the density of states (DOS) are given in  Fig.~\ref{fig:2} as described in the caption.  The less energetically advantageous structure (FM) shows flat bands and no crossing of the Fermi energy~$E_{\rm F}$ in the segments along the $z$~direction in reciprocal space (X-P  and Z-$\Gamma$ segments). The parallel orientation of spins connected by the $J_{\rm B}$ (diagonal) and $J_{\rm C}$ (along $c$) interactions causes confinement of the electrons to the $ab$~planes.  Because of the geometrical arrangement, Eu spins can only interact via conduction electrons. The effective  exchange couplings must be due to a minimization of the kinetic energy of the conduction band. Stronger confinement means higher kinetic energy. One sees that the conduction bands are flat along X-P and Z-$\Gamma$ for the FM structure, so the collinear alignment of spins confines the electrons to the $ab$ planes.  The RKKY interaction changes sign and, depending of the distance between the spins, FM or AFM alignment is favored, otherwise the conduction electrons are scattered (localized).   In all cases, the Eu $4f$ bands are well below $E_{\rm F}$ and the contribution to the conduction bands is mainly due to Co $3d$ states.

For \ecp\ we find the DOS at $E_{\rm F}$ to be between 3.5 an 4.1~states/eV\,f.u.\ (both spin directions) for the four considered magnetic structures whereas the DOS at $E_{\rm F}$ is $\approx3.2$~states/eV\,f.u.\ (both spin directions) for \bcp.  A puzzling aspect of the comparison of theory with experiment for \ecp\ is the high DOS at $E_{\rm F}$ in Table~\ref{Table:Heat_capacity} derived from the above $C_{\rm p}(T)$ measurements, which is 2.5 times larger than the calculated value of $\approx 4$~states/(eV~f.u.)\ (both spin directions) in the magnetically-ordered state for all four magnetic ordering configurations.  A similar disagreement between theory and experiment is found for the PM state of the analogous compound BaCo$_2$P$_2$, which suggests that the cause of the measured high DOS for both compounds is not interactions of the current carriers with the Eu spin helix. For all four spin configurations for \ecp\ considered, as well as for BaCo$_2$P$_2$, the DOS versus energy shows a sharp maximum (with the required height) just above $E_{\rm F}$\@. Thus, it is possible that smearing effects on the DOS due to correlations which are neglected in DFT could move part of the DOS maximum down to $E_{\rm F}$\@.  The electron-phonon interaction and many-body effects probably also contribute significant enhancements of the DOS compared with the GGA band-structure values for the two compounds. 

Regarding a comparison of the band structure of \ecp\ with that previously determined for isostructural and isoelectronic ${\rm SrCo_2P_2}$ \cite{Cuervo-Reyes2014}, the main important difference is the contribution of Eu states at $E_{\rm F}$ (about 20\% of the total DOS). For ${\rm SrCo_2P_2}$, Sr states are practically absent at $E_{\rm F}$. The Eu~$s$ and~$d$ states bridge the electronic transport along the $c$~direction and allow the establishment of a magnetic long-range order, which is absent in ${\rm SrCo_2P_2}$ \cite{Morsen1988, Teruya2014}.  Although Sr and Eu are quite similar in size and valence states, Eu is known to be more electronegative, which translates into a larger participation of its valence states.

\section{\label{Sec:Summary} Summary}

The structural refinement of trigonal \ecp\ confirms previous results \cite{Marchand1978}.  The $ab$-plane $\rho(T)$ data show that \ecp\ is metallic.  These data also show a sharp increase in slope upon cooling below $T_{\rm N} = 66$~K, in agreement with a previous report \cite{Nakama2010}.  The magnetoresistance at $H=10$~T is negligible over the measured temperature range 1.8~K~$\leq T\leq 300$~K\@.

The magnetic contribution $C_{\rm mag}(T)$ to the measured $C_{\rm p}(T)$ is extracted using our measured $C_{\rm p}(T)$ for \bcp\ as a reference compound.  We find that $C_{\rm mag}(T)$ for $T<T_{\rm N}$ is in good agreement with the prediction of MFT\@.  A small contribution to $C_{\rm mag}(T)$ above~$T_{\rm N}$ from dynamic short-range AFM order is found up to $\sim 100$~K, amounting to only $\approx 7$\% of the disordered entropy $R\ln(2S+1)$.  $C_{\rm p}(T,H)$ measurements in the ranges 1.8~K~$\leq T\leq$~80~K and 0~T~$\leq H\leq 7$~T exhibit a reduction in $T_{\rm N}$ with increasing~$H$ that agrees with the MFT prediction.  The density of states at the Fermi energy of \ecp\ and \bcp\ are found from the heat capacity data to be large, 10 and 16~states/eV per formula unit for \ecp\ and \bcp, respectively.  These values are enhanced by a factor of $\sim2.5$ compared to those found from our DFT band-structure calculations.

After correcting for the influences of sample-shape, magnetic-dipole and single-ion anisotropies, the $\chi_{ab}(T\leq T_{\rm N})$ arising from Heisenberg interactions is well fitted by our MFT with the helix pitch close to that found from neutron diffraction measurements \cite{Reehuis1992}.  The $M_c(H)$ data at 2~K are nearly linear in field up to $H=14$~T, where $H$ is parallel to the helix axis, again in good agreement with the MFT for a helical AFM structure.  The extrapolated critical field for the second-order transition from the canted AFM state to the PM state is estimated as $H_{\rm c}\sim 28$~T, in approximate agreement with the extrapolated value $H_{\rm c}\sim 25$~T obtained from the $C_{\rm p}(H,T)$ measurements up to $H=7$~T\@.  With the field in the $ab$~plane of the ordered moments, a metamagnetic transition was found at $H \approx 7$~T at $T=2$~K with structure suggesting spin-reorientation transitions occur with increasing field as previously predicted for a helix \cite{Nagamiya1967, Kitano1964}.  The metamagnetic transitions occur at decreasing $H$ with increasing~$T$\@.  It would be interesting to further investigate the field dependence of the AFM structure.

The Heisenberg exchange interactions in \ecp\ within the $J_0$-$J_{z1}$-$J_{z2}$ MFT model in Fig.~\ref{Fig:J0_Jz1_Jz2_model_helix} are obtained.  The dominant Eu--Eu exchange interactions $J_0$ are within the $ab$~plane Eu layers and are FM, consistent with expectation from the positive Weiss temperature $\theta_{\rm p}$.  The interactions $J_{z1}$ and $J_{z2}$ between Eu spins in nearest- and next-nearest-layers are both AFM\@.  From these interactions, we estimated the exchange interactions between first-, second- and third-nearest-neighbor Eu spins.  The signs of these interactions are confirmed from our DFT calculations.  We also obtain an estimate of the classical ground-state energy of the helix from the MFT\@.  The spin-wave spectrum was studied and the low-energy spin-wave excitations of the helix calculated.  From this information, we obtain the coefficient of the spin-wave $T^3$ contribution to the low-temperature $C_{\rm p}(T)$ and find that it is comparable with the lattice contribution.

Thus the experimental $C_{\rm mag}(T\leq T_{\rm N})$, $\chi_{ab}(T\leq T_{\rm N})$, $\chi_{c}(T\leq T_{\rm N})$ and $M_c(H,T=2~{\rm K})$ data for \ecp\ crystals are all in good agreement with the corresponding predictions of MFT, where the pitch of the helix needed to fit the $\chi_{ab}(T\leq T_{\rm N})$ data agrees well with the result obtained from neutron diffraction measurements \cite{Reehuis1992}.  DFT electronic structure calculations confirm the FM Eu--Eu interactions within the $ab$~plane and the AFM interactions between Eu spins in adjacent Eu layers and in next-nearest Eu layers as inferred from MFT\@.  Thus we conclude that \ecp\ is a model molecular-field helical Heisenberg antiferromagnet. \\

\noindent{\it Note added} --- A helical AFM ground state was recently reported for the Eu spins in the related isoelectronic and isostructural compound \eca.  The helix has a $c$-axis wave vector and $T_{\rm N} = 47$~K as determined from neutron diffraction measurements \cite{Tan2016} that are similar to those in \ecp.  This study found no observable moment on the Co atoms in the ordered state, consistent with our data and previous results \cite{Reehuis1992, Morsen1988} on \ecp.  Also reported were anisotropic $\chi$ data for single crystals of \eca\ \cite{Tan2016} resembling our data for \ecp.  These $\chi(T)$ data \cite{Tan2016} are in turn comparable to earlier $\chi(T)$ data for single-crystal \eca\ \cite{Ballinger2012}.\\

\acknowledgments

The research at Ames Laboratory was supported by the U.S. Department of Energy, Office of Basic Energy Sciences, Division of Materials Sciences and Engineering.  Ames Laboratory is operated for the U.S. Department of Energy by Iowa State University under Contract No.~DE-AC02-07CH11358.  The financial support of E.C.R.\ by SCCER Storage/Mobility is gratefully acknowledged.  E.C.R.\ also thanks the Small Molecule Crystallographic Center at ETH Z\"{u}rich for computational support.

\end{document}